# Station-keeping control of an unmanned surface vehicle exposed to current and wind disturbances

Edoardo I. Sarda, Huajin Qu, Ivan R. Bertaska and Karl D. von Ellenrieder

*Abstract*— Field trials of a 4 meter long, 180 kilogram, unmanned surface vehicle (USV) have been conducted to evaluate the performance of station-keeping heading and position controllers in an outdoor marine environment disturbed by wind and current. The USV has a twin hull configuration and a custom-designed propulsion system, which consists of two azimuthing thrusters, one for each hull. Nonlinear proportional derivative, backstepping and sliding mode feedback controllers were tested in winds of about 4-5 knots, with and without wind feedforward control. The controllers were tested when the longitudinal axis of the USV was aligned with the mean wind direction and when the longitudinal axis was perpendicular to the mean wind direction. It was found that the sliding mode controller performed best overall and that the addition of wind feedforward control did not significantly improve its effectiveness. However, wind feedforward control did substantially improve the performance of the proportional derivative and backstepping controllers when the mean wind direction was perpendicular to the longitudinal axis of the USV. An analysis of the length scales present in the power spectrum of the turbulent speed fluctuations in the wind suggests that a single anemometer is sufficient to characterize the speed and direction of the wind acting on the USV.

## I. INTRODUCTION

Unmanned Surface Vehicles (USVs) are playing increasing roles in commercial, scientific and military applications [1]. Once equipped with advanced control systems, sensor systems, communication systems and weapon systems, they can perform a variety of missions that include sea patrol, environmental monitoring, pollutant tracking, surveillance, underwater terrain mapping and oceanographic research [2], [3], [4], [5], [6], [7], [8]. To be effective, a USV needs to be capable of autonomously performing a variety of distinct maneuvers, with trajectory tracking and station-keeping being essential in their roles. While the former is necessary to allow the vehicle to navigate within different locations, the latter allows the system to maintain constant position and heading over a period of time. A potential application of USVs is the automatic launch and recovery (ALR) of smaller unmanned systems, such as autonomous underwater vehicles (AUVs) [6], [9], [10] and object localization using acoustic [11] or vision subsystems [12], [13], [14]. Underwater object localization via acoustics can require maintaining a fixed position and orientation for up to one minute, allowing the filters in the acoustic system to remove refraction noise, thereby improving measurement accuracy [11]. The performance of the acoustic sensors can be heavily affected if the vehicle drifts during the measurement. A similar case is that of optical localization using a camera. Here, image processing algorithms may require a few seconds; however, the performance is heavily affected if the vehicle's heading is not maintained constant, since small motions of the camera may result in dramatic changes in lighting conditions and image perspective. The ALR of an AUV from a USV is a complex task that requires precise collaboration between the USV on the surface and the AUV underwater. The process can be simplified by fixing the USV position on the surface to reduce the number of moving objects, so that the problem is essentially transformed to the static docking of the AUV [15]. Thus, enabling a USV to station-keep can convert complicated tasks into simpler ones.

Since the present generation of USVs are lightweight and have relatively large windage areas, wind is a major source of disturbance during station-keeping operations [16]. While slowly varying environmental changes, such as tidal currents, can be attenuated by applying robust feedback control laws, rapid environmental changes, like the ones caused by wind, can be better counteracted by applying feedforward control theory [17]. This research highlights that the uncertain effect of currents on a small twin-hulled USV, tasked to autonomously station-keep, leads to the inability of the system to reach and maintain the desired state. It is shown that robust non-linear control theory, such as backstepping and sliding mode control, can be applied and refined for the purpose of heading and position station-keeping of a USV. It is also shown that, implementing wind feedforward control, in addition to state feedback control, allows for fast correction of the final control signal, therefore providing the appropriate control effort.

Different options for the station-keeping control of a small twin-hulled USV are presented. More precisely, three station-keeping controllers are designed and implemented on a USV: a Proportional Derivative (PD) nonlinear controller, a Backstepping Multiple Input Multiple Output (MIMO) PD nonlinear controller and a Sliding Mode MIMO nonlinear controller. A wind feedforward control feature was also designed and added to the system to assist each station-keeping feedback controller. Experimental on-water station-keeping test results are presented for all three controllers, implemented with and without the wind feedforward feature. The outcomes of this work are the development and experimental validation of an optimal station-keeping control system for a USV tasked with common objectives, such as ALR and object localization.

Here, the performance of the controllers developed in [18] and [19] is improved by using a revised and validated dynamic model for the USV16. Previously, a model of a WAM-V USV14 [20] had been adapted for the development of a station-keeping controller by simply scaling the physical parameters in the model to those of the USV16 (shown in TABLE 1 below). Apart from the propulsion system, the WAM-V USV16 is essentially a scaled version of the WAM-V USV14. This had been found to be acceptable for initial trials. However, for this effort, a more accurate model is developed for the WAM-V USV16 using the procedure outlined in [20], and through additional on-water experimentation it is found that the performance of the controllers is improved.

This paper is organized as follows. Recent advances in station-keeping control and wind feedforward control for USVs are presented in Section II. The principal characteristics of the vehicle used in this research are described in Section III, with emphasis on the propulsion system. In Section IV, the development of a control oriented state space model of the vehicle and the wind is described. Three alternatives for station-keeping control and one option for wind feedforward control are explored in Section V. The Lagrangian multiplier method with an extended thrust representation used for control allocation is explained in Section VI. In Section VII, the controllers' station-keeping performance with and without the aid of the wind feedforward controller is compared. Finally, in Section VIII, some concluding remarks are given regarding the results shown and possible future work.

## II. Literature Review

Nonlinear control of USVs is currently an active area of research, with the majority of the effort devoted towards feedback linearization and backstepping methods [21], [22], [23], [24], [25], [26], [27], as well as sliding mode control [28], [23], [27]. However, the validation of USV control laws is often limited to numerical simulation or small-scale experiments, rather than full-scale sea trials [29]. In fact, even in more technologically mature areas such as AUV control, stabilization in the presence of environmental disturbances has only been partially addressed [30]. Several solutions have been proposed for the station-keeping of surface vehicles. In [31], experiments were performed on a small underactuated USV with high windage, where a feedforward wind model was modified to accommodate a PD-based heading autopilot. In [32], the feasibility of reducing USV drift rate, considering the wave drifting effect as the vehicle is under station-keeping mode, is discussed. Elkaim and Kelbly [33] were able to add station-keeping functionality to a wind propelled autonomous catamaran for the purpose of maintaining position at a given waypoint in the presence of unknown water currents. Switching between point and orientation stabilization and discontinuous control was employed to stabilize a marine vehicle at a fixed point in the presence of a current using dipolar vector fields as guidance in [34] and [35]. Similarly, a hybrid approach was taken in [36] where multi-output PID controllers with and without acceleration feedback were used to stabilize a vehicle in high sea states by the use of an observer to estimate the peak wave frequency. The system switched to controllers better suited to handle large disturbances as the peak wave frequency estimate decreased and, correspondingly, the sea state increased. Aguiar and Pascoal [30] devised a nonlinear adaptive controller capable of station-keeping an AUV with uncertain hydrodynamic parameters in the presence of an unknown current. Backstepping also was suggested in [37] as means to station-keep a fully-actuated vehicle, although environmental disturbances were not explicitly stated in the problem formulation. Most previous work on USV station-keeping control focuses on underactuated systems and on enabling the vehicle to maintain position only, as if it was anchored, and without focusing on the USV orientation. Here we describe the development of an ideal station-keeping controller for an overactuated USV, enabling it to simultaneously maintain heading and position.

The USV application of feedforward control theory, such as wind feedforward control, still has not been widely explored. A few challenges are encountered when designing wind feedforward controllers. These include accurately measuring representative wind speed and direction and calculating the wind force coefficients. Anemometer errors can be minimized, but not eliminated, by appropriately calibrating the sensor [17]. It has been found that wind gust and turbulence can also cause large measurement errors [16]. A risk of applying wind feedforward control is that the speed and direction of the wind can vary across different parts of a vessel. Thus, especially for larger vehicles, it may not be appropriate to analyze wind effects on the whole vessel with a single point measurement. One possible solution is to use several wind anemometers to measure the wind field [16]. For small USVs, the wind acting on the vehicle can be assumed to be uniform. The placement of the anemometer on the vehicle presents another challenge. Ideally, it should be mounted such that the measurements are least affected by wind interaction with the vehicle's structure. Lastly, wind models, capable of estimating wind forces and moments acting on small marine vessels, are very limited. The vast majority of these models are based on large vessels and floating objects in the ocean, such as oil tanks, and commercial ships [16], [38], [39], [40], [41]. These models need to be validated and possibly modified for use on small vehicles.

The research presented here applies the results from previous studies to small USVs for the purpose of autonomous station-keeping of heading and position. This work also extends the results of previous studies to include the effects of combining robust control and wind feedforward control for station-keeping purposes to minimize the sources of error. This is done through extensive on-water trials.

III.     THE WAM-V USV16

*A.     Overview*

The USV used here is a 4.9m (16') Wave Adaptive Modular-Vessel (WAM-V), the USV16 (Fig. 1). It is a twin hull, pontoon style vessel designed and built by Marine Advanced Research, Inc. of Berkeley, CA USA. The vessel structure consists of two inflatable pontoons, a payload tray connected to the pontoons by two supporting arches and a suspension system. The USV16 is designed to mitigate the heave, pitch and roll response of the payload tray when the vehicle operates in waves. The vehicle's physical characteristics are shown in TABLE 1.

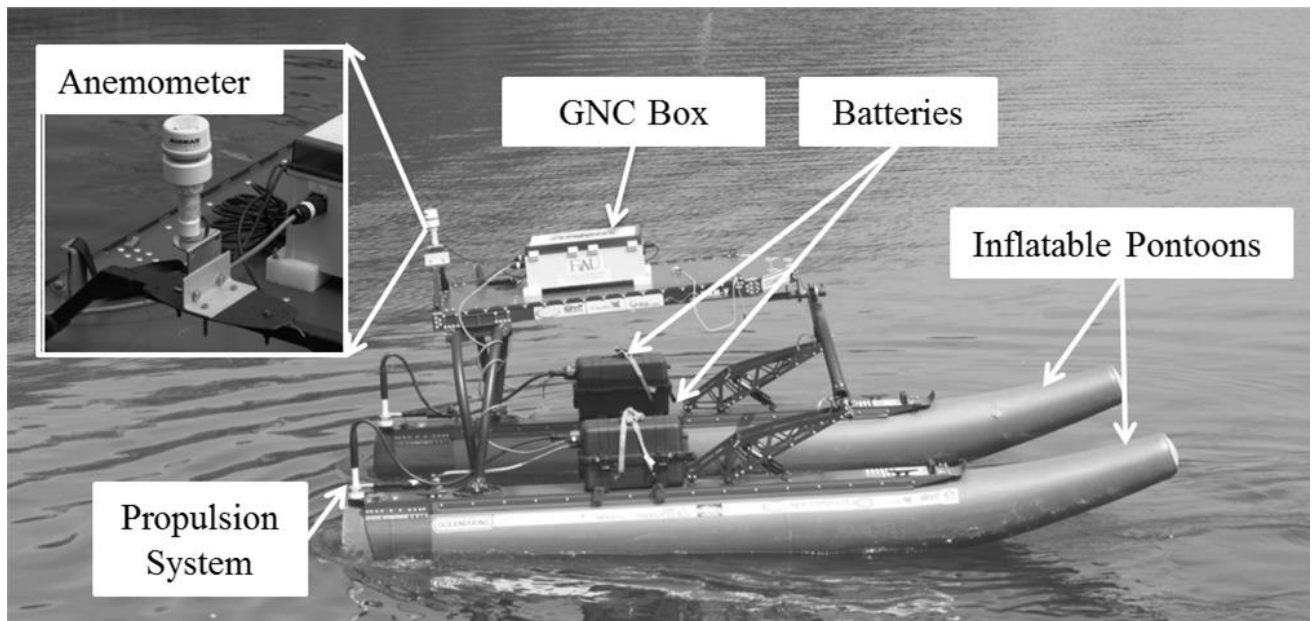

Fig. 1: The WAM-V USV16 during on-water station-keeping tests in the Intracoastal Waterway [19].

TABLE 1

PRINCIPLE CHARACTERISTICS OF THE WAM-V USV16. THE LOCATION OF THE "KEEL" IS TAKEN AS THE BOTTOM OF THE PONTOONS. W.R.T. IS AN ACRONYM FOR THE PHRASE "WITH RESPECT TO".

| Parameter | Value |
|---|---|
| Length overall (L) | 4.05 [m] |
| Length on the waterline (LWL) | 3.20 [m] |
| Draft (aft and mid-length) | 0.30 and 0.23 [m] |
| Beam overall (BOA) | 2.44 [m] |
| Beam on the waterline (BWL) | 2.39 [m] |
| Depth (keel to pontoon skid top) | 0.43 [m] |
| Area of the waterplane (AWP) | 1.6 [m$^2$] |
| Centerline-to-centerline side hull separation ($B$) | 1.83 [m] |
| Length to beam ratio (L/B) | 2.0 |
| Volumetric displacement ($\nabla$) | 0.5 [m$^3$] |
| Mass | 180 [kg] |
| Mass moment of inertia about z axis (estimated with CAD) | 250 [kg-m$^2$] |
| Longitudinal center of gravity (LCG) w.r.t. aft plane of engine pods | 1.30 [m] |

*B. Propulsion System*

In addition to its standard characteristics, a custom designed propulsion system was implemented on the USV16 (Fig. 2). This includes two 120N electric thrusters, each powered by a 12V lead acid battery, and two 160N, 15cm stroke, linear actuators capable of rotating the thrusters through an azimuthal angle of $\pm 45^o$ with respect to the vehicle's longitudinal axis (Fig. 3).

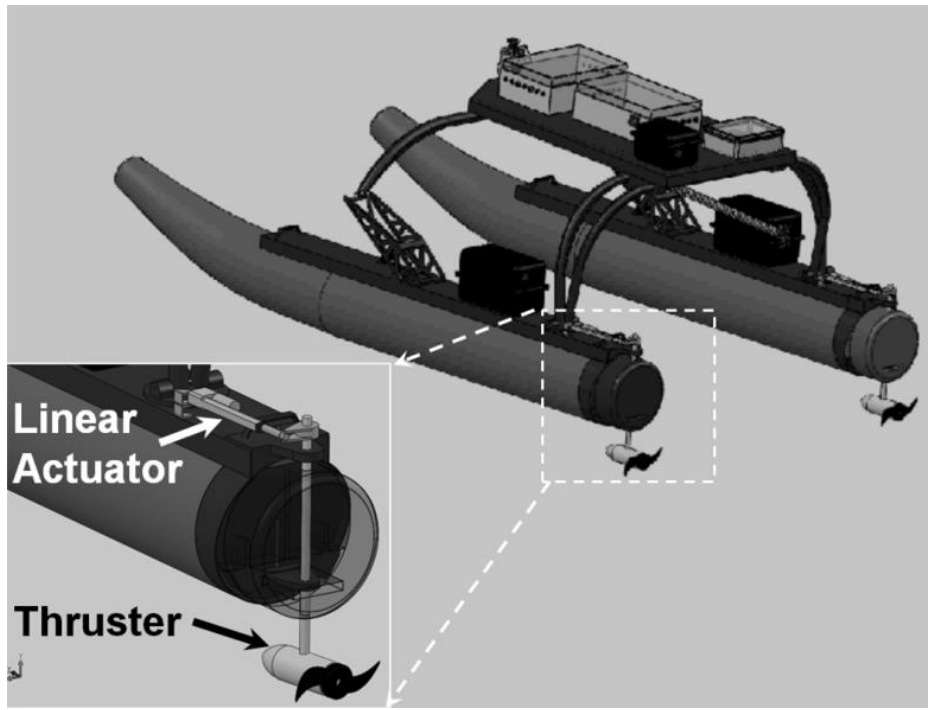

Fig. 2: WAM-V USV16 propulsion system in CAD [18].

The thrust on each pontoon ($T_p$ and $T_s$) can therefore be pivoted in various directions based on the azimuth angle on each side ($\delta_p$ and $\delta_s$), enabling the vehicle to output multiple combinations of surge and sway forces and yaw moment. Here, the subscript $s$ and $p$ stand for starboard and port sides, respectively. The moment generated by the thrusters is calculated based on the moment arms ($\boldsymbol{r}_p$ and $\boldsymbol{r}_s$), shown in Fig. 3. The configuration of the propulsion system permits the USV16 to move in surge ($u$), sway ($v$) and yaw ($r$) directions independently, so that it is an overactuated system. A system is considered overactuated when the number of actuators is greater than the number of degrees of freedom (DOF). Here, we have four actuators, namely two linear actuators and two thrusters, and three DOF ($u, v, r$).

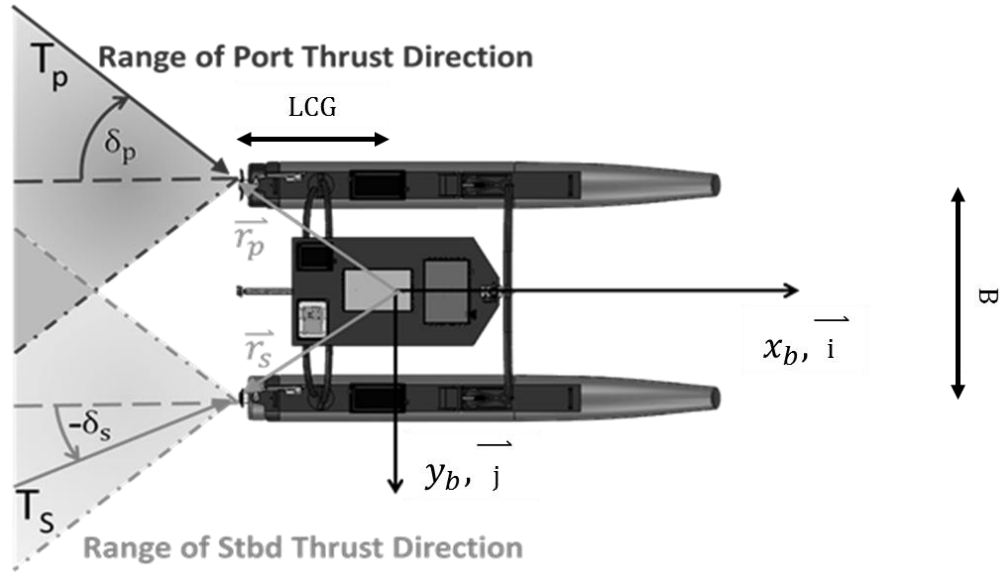

Fig. 3: Port and starboard thrust ranges directions. Both port and starboard can provide thrust at $\pm 45°$ with respect to demihull centerlines [19].

*C.     Guidance, Navigation, and Control Electronics*

A modular guidance, navigation and control (GNC) system was developed at FAU to enable USVs' autonomy [42]. The GNC system is housed in a plastic, water-proof box and contains a single-board computer (SBC), an inertial measurement unit (IMU) with global positioning system (GPS) capability, a tilt-compensated digital compass, a radio frequency (RF) transceiver, a pulse width modulation (PWM) signal generator and a custom-built printed circuit board for power distribution and communications between the SBC and instrumentation. A detailed description of the GNC system, which was implemented on the WAM-V USV16 for this work, can be found in [20]. For the purpose of this project, the key components of the GNC system are the sensor suite (IMU/GPS and digital compass), single-board computer, and RF transceiver. The IMU/GPS is an Xsens MTi-G sensor, which is used to estimate the position and orientation of the USV during operations. The GPS is Wide Area Augmentation System (WAAS) enabled and can provide up to 1 meter accuracy in both latitude and longitude, depending on cloud cover and satellite availability. The digital compass is used to monitor vehicle heading and has a resolution of 0.1 degrees. The Lightweight Communication and Data Marshaling (LCM) system [43] is utilized as the underlying architecture for the GNC software. Sensor data was transmitted using drivers incorporating the LCM system to the control architecture, and logged at 4Hz. A handheld, remote control, transmitter is used to switch between manual and on-board, computer-controlled, autonomous operation, as well as to control the thrusters and actuators when in manual mode.

An AIRMAR WeatherStation 100WX was installed on the WAM-V USV16 (Fig. 1). This ultrasonic anemometer is used to measure the apparent wind speed and direction at a sample rate of 1 Hz. The measurement data are input into the feedforward controller using an LCM process. The dynamic range of the anemometer is 0-40 m/s with a resolution of 0.1 m/s. The anemometer is located in an elevated position at the aft end of the payload tray to avoid the effects of wind blockage and interference from other structures (Fig. 1). As shown in Section V.D below, owing to the relatively small size of the USV it can be assumed that the wind speed and direction measured by a single point sensor is representative of the wind flowing past the entire vessel.

## IV. SYSTEM IDENTIFICATION

A dynamic model of the WAM-V USV16 was developed, following the same procedure as in [20]. System characterization experiments were conducted to identify the hydrodynamic coefficients to represent the WAM-V USV16 in simulation and to develop a three DOF maneuvering model of the vehicle. This model can be utilized for the design and development of various low level controllers on the WAM-V USV16. A brief description of the procedure to characterize the vehicle and the development of the state space maneuvering model is given in this section.

### A. System Characterization

A series of sea trials, including bollard pull tests, acceleration tests, circle tests, and zigzag tests, corresponding to the standard maneuvers performed on surface vessels [44], were conducted for the system identification of the WAM-V USV16. During all runs, the azimuth angles of the thrusters, $\delta_s$ and $\delta_p$ in Fig. 3, were kept at 0º. All system characterization sea trials were conducted at North Lake, Hollywood, FL. The location was chosen such that it would provide a benign environment with minimum wind, current and wave disturbances. Vehicle state, as well as wind speed and direction, were recorded throughout the experiments. The specific procedure for the tests described in this section is presented in [45]. A bollard pull test was first performed to estimate the relationship between motor commands and thrust on the vehicle. For this test, the vehicle was tied to a load scale, connected to a fixed pole. The same thrust command was given to each motor, to apply the propulsion force uniformly in the surge direction. The reading of the load scale was taken after the vehicle had put enough tension on the line to keep it steadily taut. This reading then corresponded to the total thrust the propulsion system was outputting. Five readings were recorded from the load scale, for each motor command, before being averaged. The averaged values of thrust for each motor command are shown in TABLE 2 and are plotted in Fig. 4.

TABLE 2

RELATIONSHIP OF MOTOR COMMAND AND THRUST FROM BOLLARD PULL TESTS.

| Motor Command [%] | -100 | -90 | -80 | -70 | -60 | -50 | -40 | -30 | 20 | 30 | 40 | 50 | 60 | 70 | 80 | 90 | 100 |
|---|---|---|---|---|---|---|---|---|---|---|---|---|---|---|---|---|---|
| Thrust [N] | -102 | -84 | -66 | -44 | -31 | -13 | -9 | -4 | 29 | 34 | 78 | 110 | 144 | 175 | 203 | 228 | 254 |

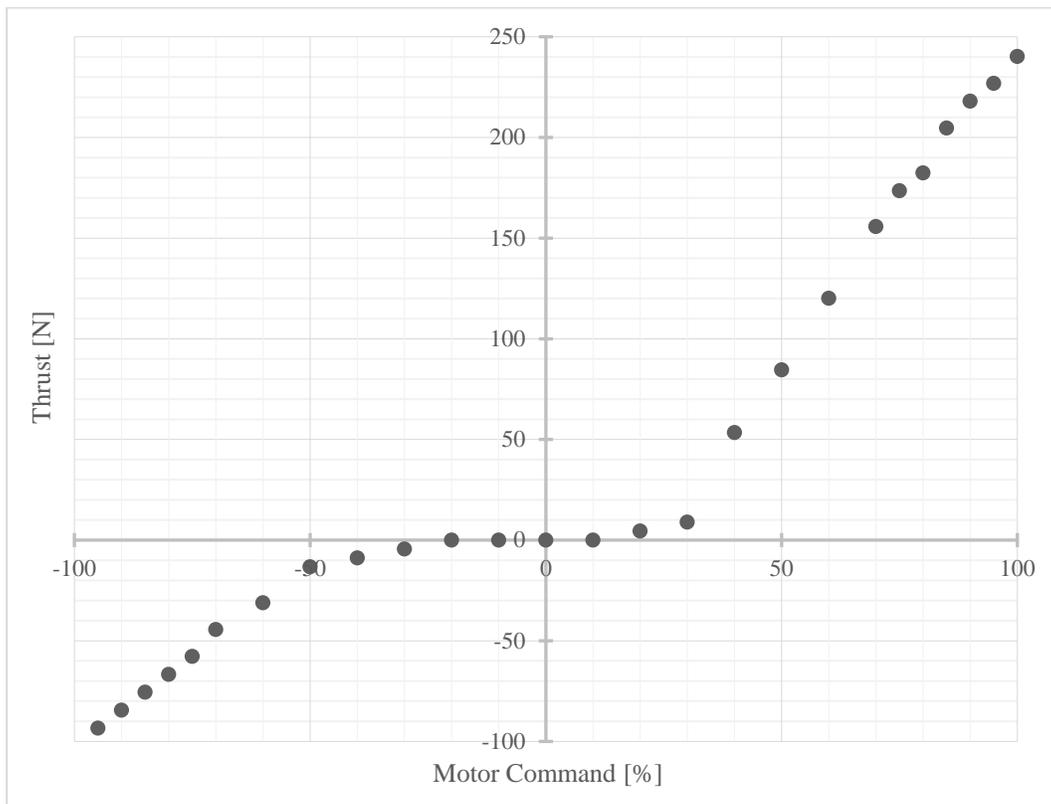

Fig. 4: Thrust vs Motor command from bollard pull test

The acceleration test was then conducted to estimate the linear and nonlinear drag terms in the surge direction. For this, the USV was started from rest and accelerated with a throttle range of 70-100% on both motors for 60 seconds. When the vehicle achieved steady-state speed, the drag forces in the surge direction were equal to propulsion forces. Linear and nonlinear drag coefficients in the surge direction, $X_u$ and $X_{u|u|}$, were determined by quadratic curve fitting of surge speeds and drag forces, as shown in Fig. 5. Note that this assumes that there is no speed dependence on the thrust developed by the propellers. This assumption does not affect the outcome of the station keeping experiments, as the forward speed of the vehicle will typically remain small during station keeping. The same assumption was previously used to develop a closed loop surge speed controller for a similar vehicle at speed and was found to work well in practice [45].

The equations used to estimate all the hydrodynamic terms needed for the dynamic model of the vehicle were chosen as in [45]. These coefficients were then linearized about a nominal surge speed of $U = 1$ m/s (the typical transit speed of the USV). Dimensional and non-dimensional terms of the equations used to estimate the hydrodynamic coefficients of the model are listed in TABLE 3.

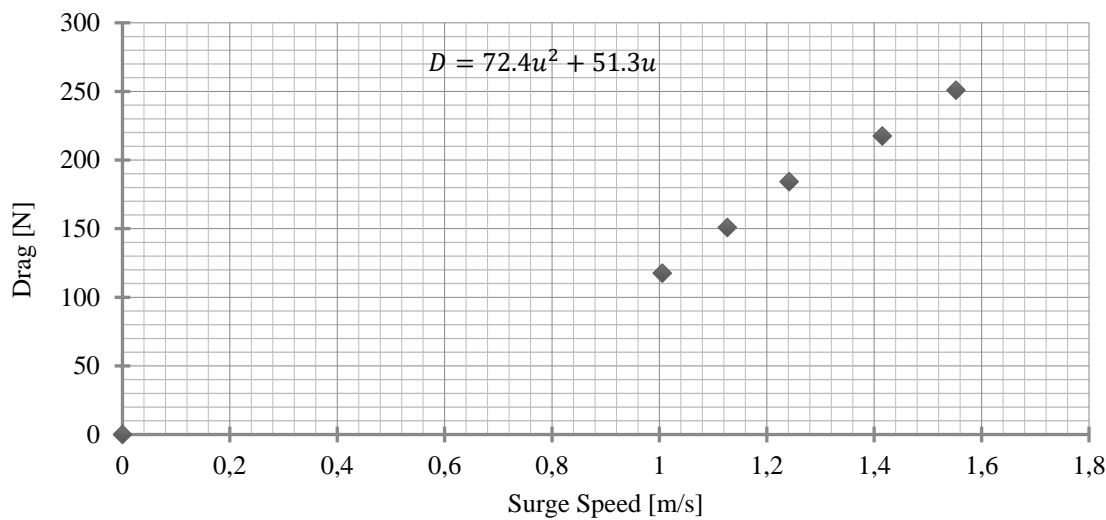

Fig. 5: Quadratic fit of surge speed and drag in surge direction for USV16 model.

TABLE 3

HYDRODYNAMIC COEFFICIENTS FOR THE WAM-V USV16. ALL HYDRODYNAMICS TERMS NOT LISTED IN THIS TABLE ARE ASSUMED TO BE ZERO. $T$ IS USED HERE TO DENOTE THE VEHICLE DRAFT, WHILE $B_{hull}$ IS THE BEAM OF THE INDIVIDUAL PONTOON HULL.

| Coefficient Name | Non-Dimensional Factor | Dimensional Term |
|---|---|---|
| $X_{\dot{u}}$ | -0.05 | $m$ |
| $Y_{\dot{v}}$ | 0.9 | $\pi \rho T^2 L$ |
| $N_{\dot{r}}$ | 1.2 | $\dfrac{4.75}{2}\pi\rho\dfrac{B}{2}T^4 + \pi\rho T^2 \dfrac{[(L-LCG)^3 + LCG^3]}{3}$ |
| $Y_{\dot{r}}$ | 0.5 | $\pi\rho T^2 \dfrac{[(L-LCG)^2 + LCG^2]}{2}$ |
| $N_{\dot{v}}$ | 0.5 | $\pi\rho T^2 \dfrac{[(L-LCG)^2 + LCG^2]}{2}$ |
| $X_u$ | | (See Fig. 5) |
| $Y_v$ | -0.5 | $\rho|v|\left[1.1 + 0.0045\dfrac{L}{T} - 0.1\dfrac{B_{hull}}{T} + 0.016\left(\dfrac{B_{hull}}{T}\right)^2\right]\left(\dfrac{\pi T L}{2}\right)$ |
| $N_r$ | $-0.65$ | $\pi\rho\sqrt{(u^2+v^2)}T^2 L^2$ |
| $Y_r$ | -0.4 | $\pi\rho\sqrt{(u^2+v^2)}T^2 L)$ |
| $X_{u|u|}$ | | (See Fig. 5) |

The hydrodynamic coefficients estimated with the equations shown in TABLE 3. The values obtained were then entered in the 3 DOF equations of motion, described in Section IV.B and manually tuned, so that, by applying the same input conditions in simulation and during sea trials, would result in the same state output. The objective was therefore to refine the estimated hydrodynamic coefficients, to produce a dynamic model of the USV, that could be considered representative the vehicle itself. For this purpose, the vehicle state was recorded through these sea trials and the same open-loop scenario was recreated in simulation. Manual tuning of the model was proven to be relatively easy to accomplish and sufficiently accurate for control system design. Automatic system identification of maneuvering coefficients and other system parameters along the lines of the work done in [46], [24] or [47] could be explored to further improve upon this work.

After the acceleration test, circle tests and zig-zag tests were performed. During the circle test, the vehicle was first accelerated with 100% throttle on both sides for 20 seconds, then port and starboard were set to -100% and 100% throttle respectively for 30 seconds. Following this procedure, the vehicle was able to spin in a circle around its center of gravity with minimal surge and sway velocity; the drag moment coefficient from yaw rate $N_r$ could therefore be isolated. The center of gravity was estimated using a detailed computer-aided design (CAD) model. The circle test was also performed in a different manner by setting port and starboard to 0% and 100% throttle. Following this procedure, the vehicle was able to steer around a turning radius; the drag coefficient in the sway directions from sway velocity $Y_v$ could therefore be estimated. During the zig-zag tests, the vehicle was first accelerated with 100% throttle on both sides for 20 seconds, then port and starboard were set to 100% and 0% throttle alternately four times on each side. The zig-zag test was utilized solely to evaluate the model by comparing field data with simulations. Comparison of simulation and experimental results for the acceleration test, circle test and zig-zag test is shown in Fig. 6, Fig. 7 and Fig. 8 respectively. In Fig. 6, it can be noted that the theoretical vehicle deceleration for both 70% and 100% throttle commands was more rapid than the experimental decelleration under the same cirumstances. A possible explanation is that the water accelerated by the USV was still moving forward when conductiong the experiments, therefore the vehicle began to decelerate, leading to a slower decay rate. In addition, unmodelled wind and current may have played a major role is affecting the decelleration of the USV during sea trials.

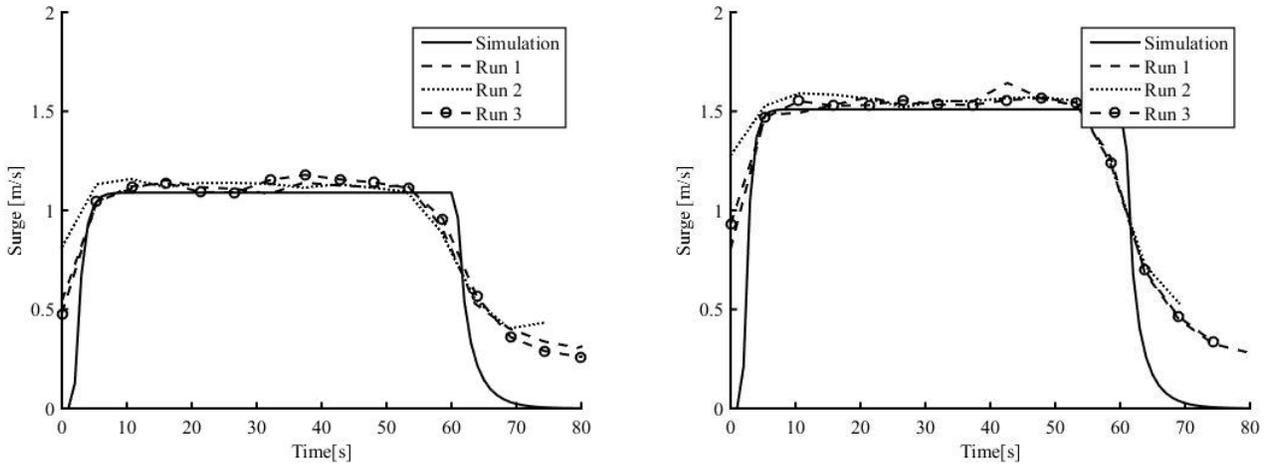

Fig. 6: Surge speed of simulation and experimental results of 70% throttle command (left) and 100% throttle command (right) during acceleration tests.

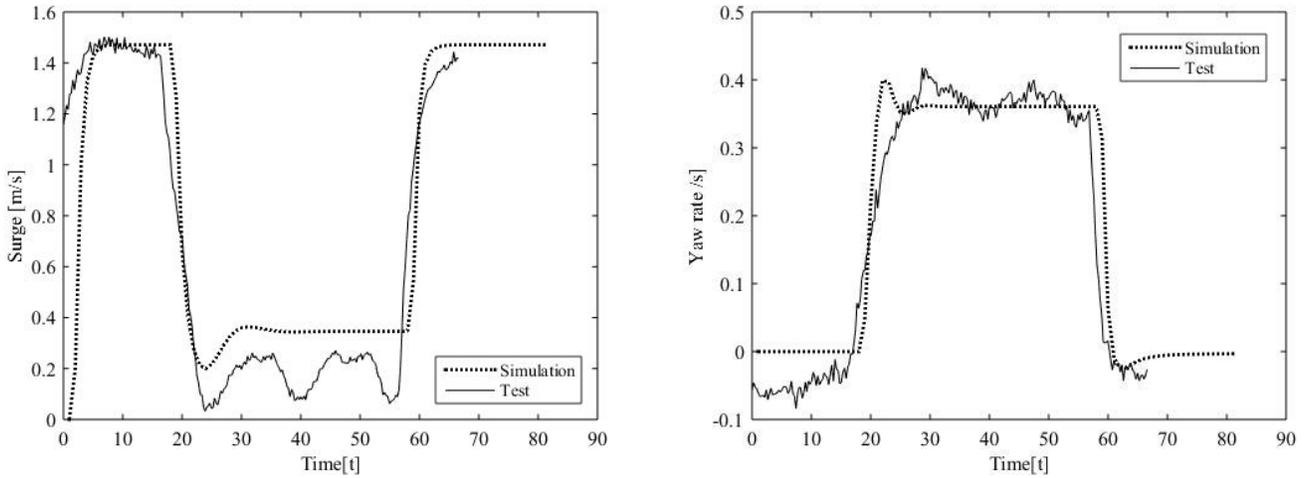

Fig. 7: Surge and yaw speeds of simulation and experimental results during circle tests.

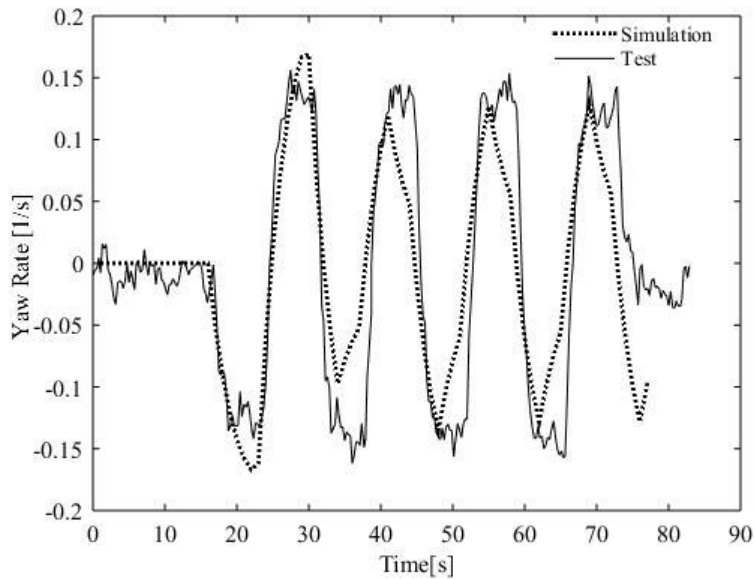

Fig. 8: Yaw rates of simulation and experimental results during zigzag tests with 100% throttle command.

It can be noted that the estimation of the hydrodynamic coefficients has been done at relatively high speeds, but the target application is station keeping under wind perturbations. The reason for this is that the effects of environmental disturbance, combined with uneven forcing on the vehicle (e.g. slightly different thrust on each side or center of gravity slightly off its desired location due to uneven weight distribution), become difficult to manage in open loop control to generate standard systems identification maneuvers at low speeds. Thus, estimation of the hydrodynamic coefficients was performed for a range of speeds between ~0.7 m/s and ~1.5 m/s (which is the maximum speed achievable with our propulsion system in no wind).

*B.    Equations of Motion*

Here, the WAM-V USV16 has been configured to operate at low speeds (1-2 knots) and in mild sea states (SS 0/1). Thus, the vehicle's motion is assumed to be planar with linear motion in the $x_b$ and $y_b$ directions and rotation about the $z_b$ axis (Fig. 9 [18]). The model designates the body-fixed origin at the center of gravity and assumes port/starboard symmetry, making $x_G = y_G = 0$. A three DOF (surge, sway and yaw) dynamic model is used to develop the equations of motion:

$$\boldsymbol{M}\dot{\boldsymbol{v}} + \boldsymbol{C}(\boldsymbol{v})\boldsymbol{v} + \boldsymbol{D}(\boldsymbol{v})\boldsymbol{v} = \boldsymbol{\tau} + \boldsymbol{\tau_w}, \tag{1}$$

and

$$\dot{\boldsymbol{\eta}} = \boldsymbol{J}(\boldsymbol{\eta})\boldsymbol{v}. \tag{2}$$

Where $\boldsymbol{M}$ is the mass matrix, $\boldsymbol{C}(\boldsymbol{v})$ is the Coriolis matrix, $\boldsymbol{D}(\boldsymbol{v})$ is the drag matrix, $\boldsymbol{\tau}$ is the vector of forces and moment generated by the propulsion system, and $\boldsymbol{\tau_w}$ is the vector of forces and moment caused by the wind. $\boldsymbol{M}$ and $\boldsymbol{C}(\boldsymbol{v})$ include both rigid body terms ($\boldsymbol{M_{RB}}$ and $\boldsymbol{C_{RB}}(\boldsymbol{v})$) and added mass terms ($\boldsymbol{M_{AM}}$ and $\boldsymbol{C_{AM}}(\boldsymbol{v})$). $\boldsymbol{D}(\boldsymbol{v})$ includes both the linear drag term ($\boldsymbol{D_l}$) and nonlinear drag term ($\boldsymbol{D_{nl}}(\boldsymbol{v})$). The vector $\dot{\boldsymbol{\eta}}$ describes the vehicle's North ($\dot{x}$), East ($\dot{y}$) velocities and the angular velocity ($\dot{\psi}$) around the $z$ axis in an inertial reference frame, $\boldsymbol{\eta} = [x, y, \psi]^T$, and the vector $\boldsymbol{v}$ contains the vehicle surge velocity ($u$), sway velocity ($v$) and yaw rate ($r$) in the body-fixed frame, thus $\boldsymbol{v} = [u\ v\ w]$ and $\dot{\boldsymbol{\eta}} = [\dot{x}\ \dot{y}\ \dot{\psi}]$. These two coordinate systems are illustrated in Fig. 9 [18].

The transformation matrix for converting from body-fixed to earth-fixed frames is given by the rotation matrix:

$$\boldsymbol{J}(\boldsymbol{\eta}) = \begin{bmatrix} \cos\psi & -\sin\psi & 0 \\ \sin\psi & \cos\psi & 0 \\ 0 & 0 & 1 \end{bmatrix}. \tag{3}$$

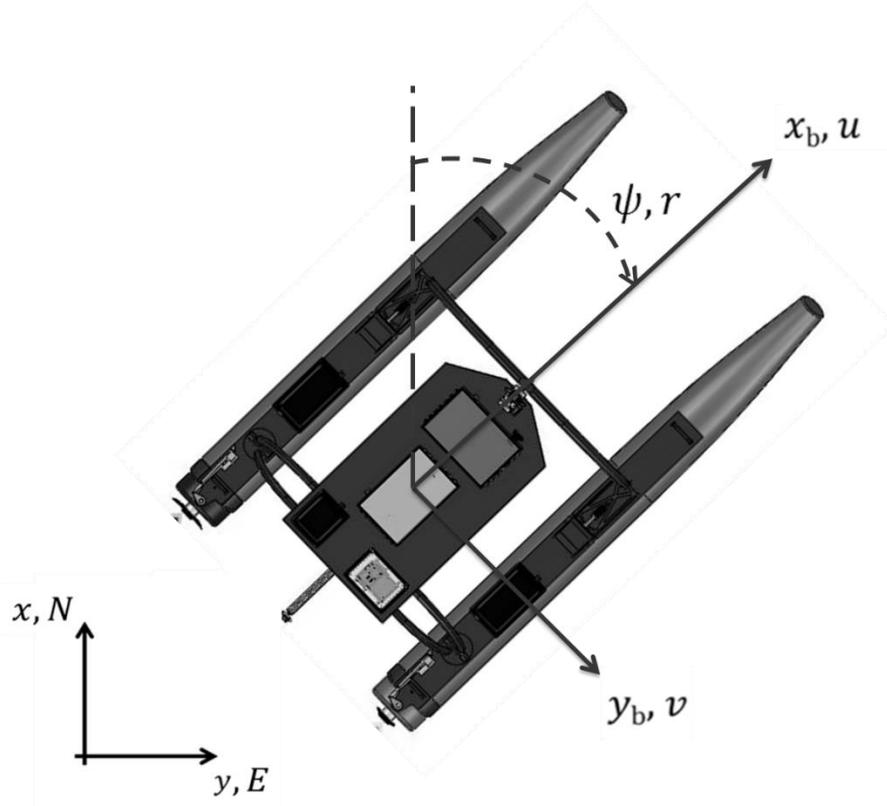

Fig. 9: Top view of WAM-V USV16 with body-fixed coordinate system overlaid. $x_b$ and $y_b$ denote vessel surge and sway axes, respectively.

$\boldsymbol{M}$ is an inertia tensor that is the sum of a rigid body mass matrix, $\boldsymbol{M_{RB}}$, and an added mass matrix, $\boldsymbol{M_{AM}}$:

$$\boldsymbol{M} = \boldsymbol{M_{RB}} + \boldsymbol{M}_{AM} = \begin{bmatrix} m - X_{\dot{u}} & 0 & -my_G \\ 0 & m - Y_{\dot{v}} & mx_G - Y_{\dot{r}} \\ -my_G & mx_G - N_{\dot{v}} & I_z - N_{\dot{r}} \end{bmatrix}, \tag{4}$$

where, $m$ denotes the mass of the USV16, $x_G$ and $y_G$ are the coordinates of the vessel center of gravity in the body-fixed frame, and $I_z$ denotes moment of inertia about the $z_b$-axis. All the terms representing the hydrodynamic coefficients in the mass matrix utilize SNAME (1950) [41] nomenclature for representing a force or moment created by motion in a specific degree of freedom. The subscript on each coefficient denotes the cause of the force/moment (e.g $Y_{\dot{r}}$ prodcues a force in the $y_b$ direction from a change in the yaw rate $\dot{r}$).

$\boldsymbol{C(v)}$ is a Coriolis matrix, which includes the sum of a rigid body term, $\boldsymbol{C_{RB}(v)}$, and added mass term, $\boldsymbol{C_{AM}(v)}$:

$$\boldsymbol{C(v)} = \boldsymbol{C_{RB}} + \boldsymbol{C_{AM}} = \begin{bmatrix} 0 & 0 & -m(x_G r + v) \\ 0 & 0 & -m(y_G r - u) \\ m(x_G r + v) & m(y_G r - u) & 0 \end{bmatrix}$$
$$+ \begin{bmatrix} 0 & 0 & Y_{\dot{v}} v + \left(\dfrac{Y_{\dot{r}} + N_{\dot{v}}}{2}\right) r \\ 0 & 0 & -X_{\dot{u}} u \\ -Y_{\dot{v}} v - \left(\dfrac{Y_{\dot{r}} + N_{\dot{v}}}{2}\right) r & +X_{\dot{u}} u & 0 \end{bmatrix} \tag{5}$$

$$\boldsymbol{D}(\boldsymbol{v}) = \boldsymbol{D}_l + \boldsymbol{D}_{nl}(\boldsymbol{v}) \tag{6}$$

where,

$$\boldsymbol{D}_l = \begin{bmatrix} X_u & 0 & 0 \\ 0 & Y_v & Y_r \\ 0 & N_v & N_r \end{bmatrix} \tag{7}$$

and,

$$\boldsymbol{D}_{nl} = \begin{bmatrix} X_{u|u|}|u| & 0 & 0 \\ 0 & Y_{v|v|}|v| + Y_{v|r|}|r| & Y_{r|v|}|v| + Y_{r|r|}|r| \\ 0 & N_{v|v|}|v| + N_{v|r|}|r| & N_{r|v|}|v| + N_{r|r|}|r| \end{bmatrix}. \tag{8}$$

The surge direction port drag and starboard drag terms, $D_p$ and $D_s$, respectively, are modeled using the polynomial curve fit derived from experimental testing (Fig. 5). A coordinate transformation is carried out to obtain the velocities of each individual pontoon hull because they are offset from the CG. These transformed velocities are used in the drag model below as $u_p$ and $u_s$:

$$D_p = \left(\frac{X_{u|u|}}{2}\right)|u_p|u_p + \left(\frac{X_u}{2}\right)u_p \tag{9}$$

$$D_s = \left(\frac{X_{u|u|}}{2}\right)|u_s|u_s + \left(\frac{X_u}{2}\right)u_s \tag{10}$$

Incorporating the moment created by the two drag forces $D_p$ and $D_s$, the term

$$\left(D_s - D_p\right)B/2 \tag{11}$$

is added to the yaw moment (not modeled in $\boldsymbol{D}(\boldsymbol{v})$).

$\boldsymbol{\tau}$ is a vector of the forces and moment generated by the propulsion system:

$$\boldsymbol{\tau} = \begin{bmatrix} T_x \\ T_y \\ M_z \end{bmatrix}. \tag{12}$$

$T_x$ , $T_y$ and $M_z$ are the thrust in $x_b$ and $y_b$ direction, and resulting moment around the $z_b$ axis. The USV16 generates the propulsion forces and turning moment with two azimuth thrusters, providing respectively $n_p$ and $n_s$ Revolutions Per Minute (RPM) on the port and starboard side respectively, and thruster turning angles $\delta = [\delta_p, \delta_s]^T$ (Fig. 3). We assume that thrust has a linear relationship with RPM, so that $T_p = \frac{n_p}{n_{max}} \times T_{max}$ and $T_s = \frac{n_s}{n_{max}} \times T_{max}$, where $n_{max}$ and $T_{max} = 120N$ are the maximum RPM and maximum thrust each thruster can output. $T_x$ , $T_y$ and $M_z$ can be calculated as in (13), (14) and (15) respectively:

$$T_x = (\boldsymbol{T_p} + \boldsymbol{T_s}) \cdot \boldsymbol{i} , \tag{13}$$

$$T_y = (\boldsymbol{T_p} + \boldsymbol{T_s}) \cdot \boldsymbol{j}, \tag{14}$$

$$M_z = \boldsymbol{r_p} \times \boldsymbol{T_p} + \boldsymbol{r_s} \times \boldsymbol{T_s}. \tag{15}$$

Here $\boldsymbol{i}$ and $\boldsymbol{j}$ are unit vectors in the $x_b$ and $y_b$ directions, respectively; $\boldsymbol{r_p} = -LCG\boldsymbol{i} - \frac{B}{2}\boldsymbol{j}$ , $\boldsymbol{r_s} = -LCG\boldsymbol{i} + \frac{B}{2}\boldsymbol{j}$ are the port and starboard moment arms, respectively; $LCG$ and $B$ are defined in TABLE 1 and shown in Fig. 3.

Lastly $\boldsymbol{\tau_w}$ denotes the forces and moment caused by wind, and is explained in the following section.

## V.  STATION-KEEPING CONTROL

Three feedback controllers have been developed and implemented on the USV16 for the same purpose, but with different formulations and characteristics. For all feedback controllers derived, the outputs are the propulsion system forces and moment $\boldsymbol{\tau}$ necessary to drive the system to its desired state, as in (1). The inputs to all of the controllers are the position and heading errors in the body-fixed coordinate frame. Since the USV16 is overactuated, a control allocation scheme must be used to apportion the output of the controller to the appropriate thrust and azimuth angle commands for the port and starboard motors (see Section VI). The block diagram in Fig. 10 shows the control flow described.

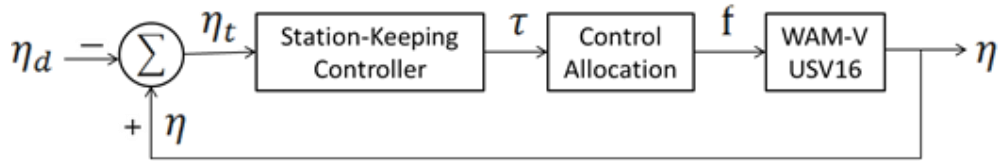

Fig. 10: Block Diagram of WAM-V USV16 Control System.

In this section, three feedback controllers and the feedforward controller used for station-keeping of heading and position on the USV16 are derived.

### A.  Proportional Derivative (PD) Control

A nonlinear Proportional Derivative (PD) controller similar to that in [36] was developed to provide a basis of comparison for the nonlinear controllers presented in the following sections. Let $\boldsymbol{\eta_d}$ be the desired pose of the vehicle in the earth-fixed frame, $\boldsymbol{\eta_d} = [x_d, y_d, \psi_d]^T$. An error vector is defined as the difference between the desired pose and the vehicle pose:

$$\boldsymbol{\eta_t} = \boldsymbol{\eta} - \boldsymbol{\eta_d}. \tag{16}$$

Using (3), this error vector can be transformed into the body-fixed frame, which leads to a nonlinear PD control law of the form

$$\boldsymbol{\tau} = -\boldsymbol{K}_P \boldsymbol{J}(\boldsymbol{\eta})^T \boldsymbol{\eta_t} - \boldsymbol{K}_D \big[ \boldsymbol{J}(\dot{\boldsymbol{\eta}})^T \boldsymbol{\eta_t} + \boldsymbol{J}(\boldsymbol{\eta})^T \dot{\boldsymbol{\eta}}_t \big]. \tag{17}$$

$\boldsymbol{K}_P$ and $\boldsymbol{K}_D$ are positive definite diagonal matrices. The gains for these matrices were first tuned in simulation and then manually refined during field experiments. The block diagram of the PD station-keeping controller described is provided in Fig. 11.

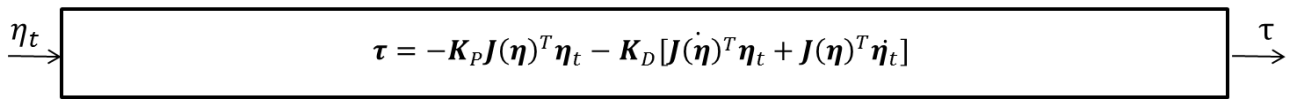

Fig. 11: PD station-keeping controller

Note that a full proportional integral derivative (PID) controller was originally implemented in simulation. However, during field trials it was found that due to the relatively low control authority of the system on the sway axis, lateral positioning errors would build up considerably. This caused the desired control forces and moments to be skewed towards regulating the lateral errors, which in turn resulted in commanded actuator configurations that were difficult to achieve, thus compromising the overall station-keeping performance. In comparison with the other non-linear controller presented in this paper, the PID station-keeping controller was unable to achieve a steady state condition. The implementation of anti-windup techniques were not sufficient to mitigate these problems. For these reasons, a PD controller, which was found to perform adequately in simulation, was used instead.

### B. Backstepping Control

Backstepping control theory was applied to station-keeping using the approach presented in [48] and [22]. Here only the final result with some necessary corrections are presented. Since the performance of the PD controller differed from that found during simulation, a MIMO backstepping controller was designed to overcome unmodeled dynamics and environmental disturbances. These two issues were considered to be the cause of the deviations between the simulated and experimental results. The implementation of a backstepping controller also circumvented possible difficulties often encountered when gain-scheduling techniques cannot be applied to linearized models. This is the case when trying to control surge and sway position $(x, y)$ and yaw angle $(\psi)$ simultaneously.

A reference trajectory is first defined as:

$$\dot{\boldsymbol{\eta}}_r = \dot{\boldsymbol{\eta}}_d - \boldsymbol{\Lambda}\boldsymbol{\eta}_t. \tag{18}$$

Here $\dot{\boldsymbol{\eta}}_d = [\dot{x}_d \quad \dot{y}_d \quad r_d]^T$ is the derivative of the desired state of the vehicle, $\boldsymbol{\Lambda}$ is a diagonal design matrix based on Lyapunov exponents, and $\boldsymbol{\eta}_t$ is the earth-fixed tracking error vector defined in (16). For the station-keeping of marine vehicles, the desired state $\boldsymbol{\eta}_d = [x_d \quad y_d \quad \psi_d]^T$ contains the desired position in the North-East-Down coordinate frame ($x_d$ and $y_d$) and desired heading $\psi_d$, while its derivative is $\dot{\boldsymbol{\eta}}_d = [0 \quad 0 \quad 0]$ since $\dot{x}_d, \dot{y}_d, r_d \rightarrow 0$ as the system approaches steady-state. Two Lyapunov functions are considered in [22] to allow the system to be feedback stabilizable, implying that a control law can be formulated to force the system to its desired state independently of its initial state. $\boldsymbol{\Lambda}$ represents the Lyapunov exponent gain matrix, which is used to allow the controller to overcome those systems dynamics neglected or overlooked in the model. The appropriate selection of $\boldsymbol{\Lambda}$ is crucial to maximize the controller performance. More specifically, large values of $\boldsymbol{\Lambda}$ may result in decent tracking performance at the cost of overlooking modelling details. On systems such as the WAM-V USV16, $\Lambda$ can be designed by imposing limits dictated by three factors:

    i.    Resonant mode: $\boldsymbol{\Lambda}$ ought to be smaller than the frequency of the lowest unmodeled resonant mode $f_r$ [49]. For the USV, this has been taken as the frequency of sloshing between the hulls in the sway direction. This frequency is

estimated as the celerity of a deep water wave [50], of wavelength equal to the BOA, divided by the BOA.

ii.   Neglected time delays: $\Lambda$ must be inversely proportional to the often-unmodeled time delays $t_u$ caused by some of the system's hardware, such as the actuators [49].

iii.  Sampling rate: similarly to i and ii, a full period of processing delay leads to $\Lambda$ being selected as a fraction of the sampling rate $f_s$ [49].

The final selection is made by choosing the value that produces the slowest decay rate of the errors. Based on i, ii, and iii, three different candidates were selected for $\Lambda$:

$$\Lambda_1 = \frac{2}{3}\pi * f_r, \tag{19}$$

$$\Lambda_2 = f_s/5 , \tag{20}$$

and

$$\Lambda_3 = \frac{1}{3 * t_u}, \tag{21}$$

where $f_r = 0.8$ Hz is the lowest resonant frequency expected, $f_s = 4$ Hz is the sampling frequency and $t_u = 2$ sec is the largest unmodeled time delay from the linear actuators that rotate the thrusters (Fig. 2). Since (21) gives the minimum bandwidth, $\Lambda$ takes the following numerical form:

$$\Lambda = \begin{bmatrix} \Lambda_3 & 0 & 0 \\ 0 & \Lambda_3 & 0 \\ 0 & 0 & \Lambda_3 \end{bmatrix} = \begin{bmatrix} 0.16 \, Hz & 0 & 0 \\ 0 & 0.16 \, Hz & 0 \\ 0 & 0 & 0.16 \, Hz \end{bmatrix}. \tag{22}$$

A measure of tracking also needs to be defined based on the Lyapunov exponents. This tracking surface is defined as:

$$\boldsymbol{s} = \dot{\boldsymbol{\eta}}_t + \Lambda\boldsymbol{\eta}_t. \tag{23}$$

The error feedback control law for station-keeping then has the following form:

$$\boldsymbol{\tau} = \boldsymbol{M_1}\big[(\boldsymbol{J}(\boldsymbol{\eta})^T\ddot{\boldsymbol{\eta}}_r + \boldsymbol{J}(\dot{\boldsymbol{\eta}})^T\dot{\boldsymbol{\eta}}_r\big] + \boldsymbol{C_1}(\boldsymbol{v})\boldsymbol{J}(\boldsymbol{\eta})^T\dot{\boldsymbol{\eta}}_r + \boldsymbol{D_1}\boldsymbol{J}(\boldsymbol{\eta})^T\dot{\boldsymbol{\eta}}_r - \boldsymbol{J}(\boldsymbol{\eta})^T K_d\boldsymbol{s} - \boldsymbol{J}(\boldsymbol{\eta})^T K_p\boldsymbol{\eta}_t. \tag{24}$$

Where $\boldsymbol{J}(\boldsymbol{\eta})$, and $\boldsymbol{\tau}$ correspond to the matrices defined in (3) and (12), respectively; $\boldsymbol{M_1}$, $\boldsymbol{C_1}(\boldsymbol{v})$, $\boldsymbol{D_1}$ are the simplified mass matrix, the Coriolis and centripetal matrix, and the drag matrix; $\boldsymbol{K_d}$ and $\boldsymbol{K_p}$ are derivative and proportional control gain matrices, respectively; $\dot{\boldsymbol{\eta}}_r$ and $\ddot{\boldsymbol{\eta}}_r$ are the derivatives of the virtual reference trajectory vector defined in (18); $\boldsymbol{\eta}_t$ is the earth-fixed tracking error vector defined in (16); and $\boldsymbol{s}$ is the tracking surface based on the Lyapunov exponent described in [48] and defined in (23).

$\boldsymbol{M_1}$, $\boldsymbol{C_1}(\boldsymbol{v})$, $\boldsymbol{D_1}$ are defined to simplify the implementation of nonlinear station-keeping controllers on the USV16. The simplification, which was also introduced in [51], [52] and [53], and then applied for the purpose of station-keeping a USV in [31], is based on three main assumptions:

i.   Negligible nonlinear drag $\boldsymbol{D_{nl}}$: During station-keeping the vehicle speed $U \to 0$, therefore $\boldsymbol{D_{nl}} \to \boldsymbol{0}$ much faster than $\boldsymbol{D_l} \to \boldsymbol{0}$. This is true when the USV speed is maintained below 1 m/s. As the vehicle's velocity increases, the effect of $D_{nl}$ becomes more important and this assumption starts to break down.

ii.   Negligible off-diagonal terms for $\boldsymbol{M}$ and $\boldsymbol{D_l}$: The effect of off-diagonal terms on the vehicle's dynamics during station-keeping is minimal compared to that of the diagonal terms. More specifically, when $U < 1$ m/s, $N_v(U^2) \ll N_r(U)$ and $Y_r(U^2) \ll Y_v(U)$ in $D_l$. Furthermore $N_v(U^2), Y_r(U^2) \to 0$ much faster than $N_r(U), Y_v(U) \to 0$. Similarly, $|m - Y_{\dot{v}}| \gg |-Y_{\dot{r}}|$ and $|I_z - N_{\dot{r}}| \gg |-N_{\dot{v}}|$ in $M$, thus $Y_{\dot{r}}$ and $N_{\dot{v}}$ are negligible at low speeds.

iii.   The terms containing $Y_{\dot{r}}$ and $N_{\dot{v}}$ in the Coriolis and centripetal matrix $\boldsymbol{C(v)}$ are negligible: A combination of approximate fore-aft symmetry and light draft suggest that the sway force arising from yaw rotation and the yaw moment induced by acceleration in the sway direction are much smaller than the inertial and added mass terms.

After applying these simplifying assumptions, $\boldsymbol{M}, \boldsymbol{C(v)}$ and $\boldsymbol{D(v)}$ in (4), (5), and (6), respectively, reduce to:

$$\boldsymbol{M_1} = \begin{bmatrix} m - X_{\dot{u}} & 0 & 0 \\ 0 & m - Y_{\dot{v}} & 0 \\ 0 & 0 & I_z - N_{\dot{r}} \end{bmatrix}, \tag{25}$$

$$\boldsymbol{C_1(v)} = \begin{bmatrix} 0 & 0 & -(m - Y_{\dot{v}})v \\ 0 & 0 & (m - X_{\dot{u}})u \\ (m - Y_{\dot{v}})v & -(m - X_{\dot{u}})u & 0 \end{bmatrix}, \tag{26}$$

$$\boldsymbol{D_1} = \begin{bmatrix} X_u & 0 & 0 \\ 0 & Y_v & 0 \\ 0 & 0 & N_r \end{bmatrix}. \tag{27}$$

The control gains are assigned in two diagonal matrices ($\boldsymbol{K_d}$ and $\boldsymbol{K_p}$) to eliminate errors in position, velocity, orientation and yaw rate. The gains were initially chosen by modeling the controller in simulation, then manually tuned during in-water tests. The block for the backstepping station-keeping controller described is provided in Fig. 12.

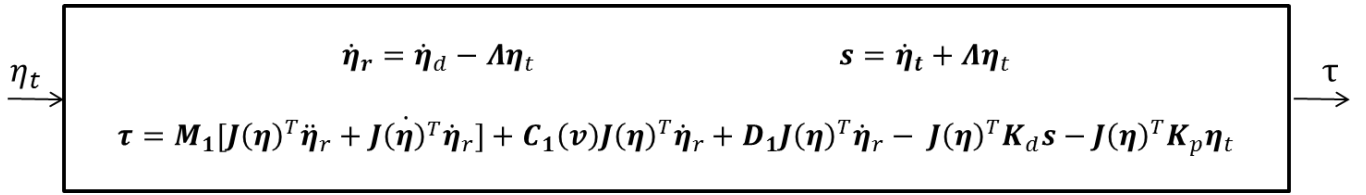

Fig. 12: Backstepping station-keeping controller.

*C.   Sliding Mode Control*

In general, robust control strategies can be implemented to improve the response of a system when its dynamic model or the nature of the environmental disturbances that act upon it are uncertain. A robust sliding mode station-keeping controller was designed and implemented on the USV16 to mitigate slowly varying environmental disturbances, such as tidal currents, that the system cannot directly measure through its sensors. Applying sliding mode control, the gains $\boldsymbol{K_d}$ and $\boldsymbol{K_p}$ in (17) and (24) are made hypothetically infinite and discontinuous, forcing the trajectories of the system to slide within a bounded sliding surface. As a result, the system diminishes its sensitivity towards parameters variations, including external disturbance and unmodeled dynamics. The advantage of this procedure is that the control signal is not required to be highly precise, since the sliding motion is invariant to small disturbances entering the system through the control channel. Sliding mode control theory was therefore considered highly suitable for the purpose of controlling a USV, tasked to maintain position and heading over an extended period of time, operating in an environment disturbed by slowly varying water currents.

A sliding surface function is first defined, similarly to (23), as:

$$s = \dot{\boldsymbol{\eta}}_t + 2\boldsymbol{\varLambda}\boldsymbol{\eta}_t + \boldsymbol{\varLambda}^2 \int_0^t \boldsymbol{\eta}_t dt. \tag{28}$$

Here $\dot{\boldsymbol{\eta}}_t$, $\boldsymbol{\eta}_t$, $\boldsymbol{\varLambda}$ are the same terms already defined in Section V.B, however an integral term is added to account for slowly varying sources of errors. This shall enable the system to be more stable over extended periods of time, at the cost of requiring a longer period of time to reach steady-state. The robustness of the controller is therefore prioritized over its performance. The same integral term is also introduced, for the purpose of minimizing unmodeled environmental disturbance, in a newly defined reference trajectory:

$$\dot{\boldsymbol{\eta}}_r = \dot{\boldsymbol{\eta}}_d - 2\boldsymbol{\varLambda}\boldsymbol{\eta}_t - \boldsymbol{\varLambda}^2 \int_0^t \boldsymbol{\eta}_t dt. \tag{29}$$

Where $\dot{\boldsymbol{\eta}}_d$ is the derivative of the desired state as defined in the Section V.B. In order to prevent integral error accumulation in the sway direction, which had been observed in simulations, the value of the integral sway error term $\int_0^t \boldsymbol{\eta}_t dt$ in equations (28) and (29) was multiplied by 0.1 when the control output $\boldsymbol{\tau}$ was saturated. In this way, scaling of the integral portion of the control signal after saturation worked as an anti-windup technique to prevent the continuous saturation of the control signal. The control law then ensures that if the system deviates from the surface defined in (28), it is forced back to it. Once on the surface, the under-modeled system reduces to an exponentially stable, second-order system. The system's response therefore depends heavily on the choice of the sliding surface. As a result the system will possess considerable robustness against slow varying external perturbations, like currents, and incorrectly modeled dynamics. The sliding mode control law is similar to (24) and is defined as follows:

$$\boldsymbol{\tau} = \boldsymbol{M}_1\big[(\boldsymbol{J(\eta)}^T\ddot{\boldsymbol{\eta}}_r + \boldsymbol{J(\dot{\eta})}^T\dot{\boldsymbol{\eta}}_r\big] + \boldsymbol{C}_1(v)\boldsymbol{J(\eta)}^T\dot{\boldsymbol{\eta}}_r + \boldsymbol{D}_1(v)\boldsymbol{J(\eta)}^T\dot{\boldsymbol{\eta}}_r - \boldsymbol{J(\eta)}^T\boldsymbol{R} * sat(\boldsymbol{E}^{-1} * \boldsymbol{s}). \tag{30}$$

It can be noted that the only difference between (24) and (30) is the last term. For the case of sliding mode control, the last term in the control law includes the bound on the uncertainties $\boldsymbol{R}$ and the boundary layer thickness $\boldsymbol{E}$ around the sliding surface $\boldsymbol{s}$, in place of the gains $\boldsymbol{K}_d$ and $\boldsymbol{K}_p$ in the backstepping control law, defined in (24). In (30), $\boldsymbol{R}$ can be considered a positive definite diagonal gain matrix and $\boldsymbol{E}$ is a vector defining the boundary layer within which the system will slide along the surface $\boldsymbol{s}$.

The saturation function ($sat$) is used instead of the traditional signum function to reduce the chattering effect, as suggested in [48]. In a situation where the system is continuously disturbed by external forces that make it deviate from steady state, applying a discontinuous control signal enables it to deviate momentarily from its standard behavior. As a result, the system is forced back to its desired state. This is exactly what happens when attempting to station-keep a USV in an environment disturbed by slowly varying, unsteady, and unmodeled currents. It will be shown in Section VII, that when the disturbance forces the vehicle away from its desired state, the sliding mode controller is able to bring the system back to its desired state, by exiting the sliding surface bounded by $E$ momentarily. As the system reenters the boundary layer it will constantly slide along the surface minimizing the error.

In (30), the bound on the uncertainties $R$ acts similarly to controller gains $K_d$ and $K_p$ in (24). The difference is that, once the system enters the boundary layer $E$, the discontinuous control signal forces the system to slide along a cross-section of the state space $s$, bounded by $R$. In other words, as the system's errors are within specific boundaries dictated by $E$, the control signal will vary based on $s$, so that $s < |R|$. A proper representation of such phenomena requires a three-dimensional plot to show the sliding surface bounded by $E$ and $R$. For simplicity, an illustration of a linear signal for each element in $R$ and $E$ is given in a two-dimensional plot and can be seen in Fig. 13.

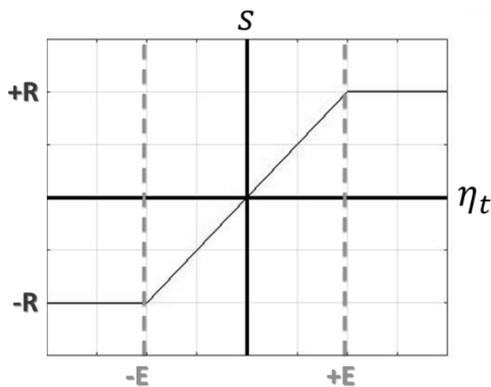

Fig. 13: Illustration of Saturation Function for a single saturation argument for sliding mode controller.

Both, the bound on the uncertainties matrix $R$ and the boundary layer thickness $E$, were initially selected based on the results obtained testing the PD and the backstepping controllers previously described, then tuned during in-water testing. More precisely, the values of all the terms in $E$ were chosen to be the average steady-state error resulting from on-water testing of the backstepping station-keeping controller described in the Section V.A and V.B, while $R$ was treated as a gain matrix, and therefore, it was iteratively tuned during experimental trials of the sliding mode controller described. It is important to note that that chattering and saturation of the control signal can compromise the functionality of the sliding mode controller, when used for the purpose of station-keeping heading and position of a USV. This was evident during sea trials. Fine tuning the values of $R$ and $E$ was therefore crucial to enable the vehicle to perform its desired maneuver. A systematic procedure was therefore developed for tuning the sliding mode controller:

i. Set the values of all terms in $E$ to be the steady state errors observed in the past.

ii. Increase all terms in $R$ until the system is successfully driven to steady state.

iii.    If any of the controller outputs seem to be constantly saturated, increase the corresponding term in $\boldsymbol{E}$.

iv.    If any of the errors seem to be too large, decrease the corresponding value of $\boldsymbol{E}$.

Note that it is very common to associate large errors with small values of $\boldsymbol{R}$, when in reality these errors may be due to inaccurate tuning of the boundary layer thickness $\boldsymbol{E}$, or vice versa. For this reason, sometimes it is best to restart the tuning procedure from scratch. In general, it is essential to remember that $\boldsymbol{R}$ ensures the system reaches the desired state with a certain margin of error, which is dictated by $\boldsymbol{E}$. The control block for the sliding mode station-keeping controller described is provided in Fig. 14.

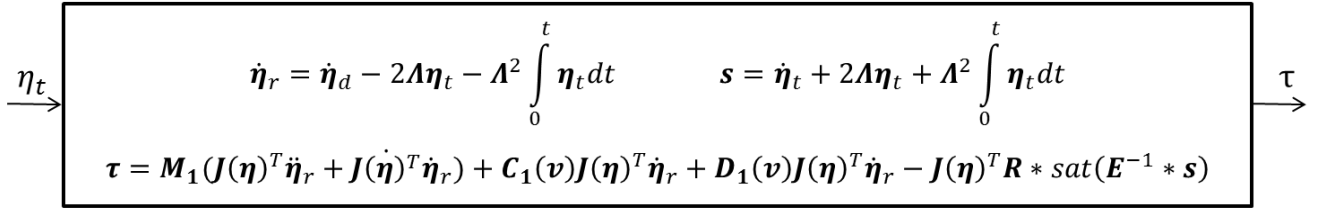

Fig. 14: Sliding Mode station-keeping controller.

*D.    Wind Feedforward Control*

The wind forces and moment acting on the vehicle are modeled as:

$$\boldsymbol{\tau_w} = q \begin{bmatrix} C_X(\gamma_{rw})A_{Fw} \\ C_Y(\gamma_{rw})A_{Lw} \\ C_N(\gamma_{rw})A_{Lw}L_{aa} \end{bmatrix}, \tag{31}$$

where $A_{Fw}$ and $A_{Lw}$ are the frontal and lateral projected windage area (Fig. 15), $\gamma_{rw}$ is apparent angle of attack [17] and $q$ is the dynamic pressure found using

$$q = \frac{1}{2}\rho_a V_{rw}^2, \tag{32}$$

where $\rho_a = 1.2$ kg/m$^3$ is the density of the air, and $V_{rw}$ is the apparent wind speed. Both apparent wind speed and direction were measured from the USV when stationary with the ultrasonic anemometer described in Section III. Representative wind data, including apparent wind speed and direction are shown in Fig. 16. The corresponding power spectral density $S(f)$ of the turbulent speed fluctuations in the wind data in Fig. 16, normalized by the intensity of the turbulent kinetic energy $TKE$ (33), where $\bar{V}_{rw}$ is the average wind speed during the measurement period [54], is shown in Fig. 17. The turbulent speed fluctuations are calculated using $V'_{rw} \equiv (V_{rw} - \bar{V}_{rw})$. The intensity of the turbulence is defined as $s = TKE/\bar{V}_{rw}$.

$$\text{TKE} \equiv \sqrt{\overline{V_{rw}'^2}} \tag{33}$$

An examination of the wind power spectrum in Fig. 17 shows that about 90 percent of the turbulent kinetic energy occurs at frequencies less than about $f = 0.03$ Hz. Using the average wind speed $\bar{V}_{rw} = 2.43$ m/s observed during these measurements, the corresponding length scales of the turbulent fluctuations in the wind are expected to be on the order of $L_t \sim \bar{V}_{rw}/f = 78$ m and larger. As the length of the USV16 is much smaller than $L_t$, a single point measurment of the wind speed and direction is taken to be sufficiently representative of the wind characteristics acting across the entire vehicle.

$C_X, C_Y, C_Z$ are the wind coefficients for the surge, sway and yaw axes. In [40], the wind coefficients for merchant ships based on 8 parameters were estimated, from wind loads on very large crude carriers (150 000 to 500 000 DWT class) from [55] and [56]. The non-dimensionalized wind coefficients could then be computed as a function of $\gamma_{rw}$. For the purpose of this research, the method advised in [57] for ships that are symmetrical with respect to the xz planes was adopted for the USV16.

The wind coefficients for horizontal plane motions can be approximated by:

$$C_X(\gamma_{rw}) = -c_x cos(\gamma_{rw}), \tag{34}$$

$$C_Y(\gamma_{rw}) = c_y sin(\gamma_{rw}), \tag{35}$$

$$C_Z(\gamma_{rw}) = c_z sin(2\gamma_{rw}), \tag{36}$$

In [57], it is suggested that $0.5 \le c_x \le 0.90$, $0.7 \le c_y \le 0.95$, and $0.05 \le c_z \le 0.20$. These values were therefore initially set to $c_x = 0.70$, $c_y = 0.80$, $c_z = 0.10$, and then manually tuned based on the vehicle performance, leading to final values of $c_x = 0.50$, $c_y = 0.50$, $c_z = 0.33$.

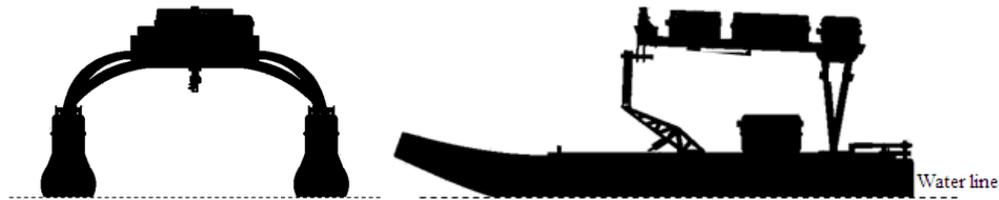

Fig. 15: Frontal (left) and lateral (right) projected areas [19].

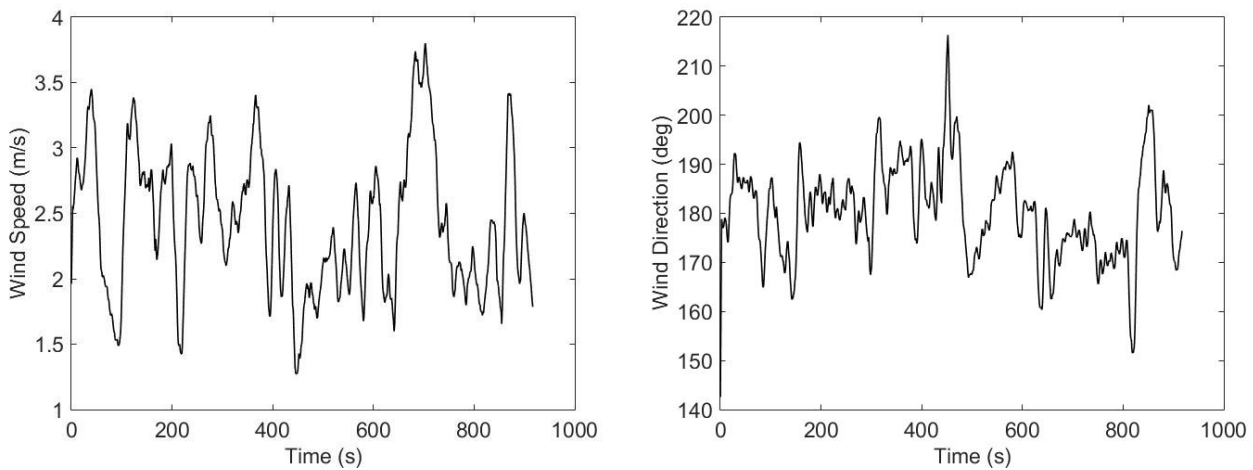

Fig. 16: Apparent wind speed (left) and direction (right), collected from a stationary vehicle.

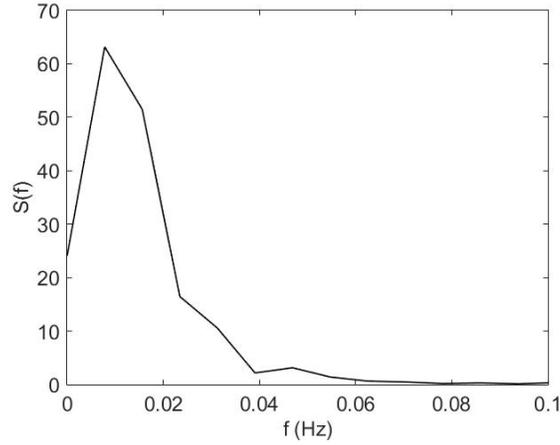

Fig. 17: Normalized power spectral density of the turbulent speed fluctations in the wind data recorded in Fig. 16.

The apparent wind speed $V_{rw}$ and angle of attack $\gamma_{rw}$ could also be estimated given the true wind speed $V_w$ and direction $\beta_w$. This is often done when a weather station on land is used as primary sensor to estimate the wind disturbance [58], [59]. Since weather stations are rigidly fixed at a specific location on land, they are able to continuously measure true wind speed $V_w$ and direction $\beta_w$. However, this approach requires the major assumption that the characteristics of the wind at the weather station can be considered representative of the wind acting on the vehicle. This is only acceptable if the USV is operating within a short distance from the weather station on land. $V_w$ can then be decomposed in true wind velocity in each direction:

$$u_w = V_w \cos(\beta_w - \psi), \tag{37}$$

$$v_w = V_w \sin(\beta_w - \psi). \tag{38}$$

Here, $u_w$ is the true wind speed in the $x_b$ direction, $v_w$ is the true wind speed in the $y_b$ direction. The true wind direction $\beta_w$ can also be calculated:

$$\beta_w = \gamma_w + \psi. \tag{39}$$

Here $\gamma_w$ is the true wind angle of attack and $\psi$ is the vehicle's heading. Knowing the vehicle's surge ($u$) and sway ($v$) velocities from (2), the relative wind speed in the body-fixed coordinate frame in each direction, $u_{rw}$ and $v_{rw}$, can be calculated:

$$u_{rw} = u_w - u, \tag{40}$$

$$v_{rw} = v_w - v. \tag{41}$$

Finally, the apparent wind speed $V_{rw}$ and angle of attack $\gamma_{rw}$ in the body-fixed coordinate frame are obtained as follows:

$$V_{rw} = \sqrt{u_{rw}^2 + v_{rw}^2}, \tag{42}$$

$$\gamma_{rw} = -\tan^{-1}(v_{rw}/u_{rw}). \tag{43}$$

Thus, the wind forces and moment acting on the vehicle can be derived as long as either true or apparent wind speed and direction are given. For station-keeping maneuvers, the vehicle velocity $\boldsymbol{U} \to 0$ as the system reaches steady-state, therefore it is assumed that $u = v = 0$, leading to $V_{rw} = V_w$ and $\gamma_{rw} = \gamma_w$ at all times.

Based on the dynamic model of the USV16 and wind model explained above, a wind feedforward controller was designed to mitigate the wind disturbance. In similar studies, it has been noted that the wind feedforward controller should be used with caution because it has the potential to make the system unstable or reduce the performance of the main feedback controller [16]. When the feedforward controller is implemented, the scheme shown in Fig. 10 is modified to accommodate an additional control input $-\boldsymbol{\tau_w}$ that acts to oppose the wind forces and moment acting on the vehicle. $\boldsymbol{\tau_w}$ is calculated based on the wind model in (31). The updated block diagram, which includes the wind feedforward controller, can be seen in Fig. 19.

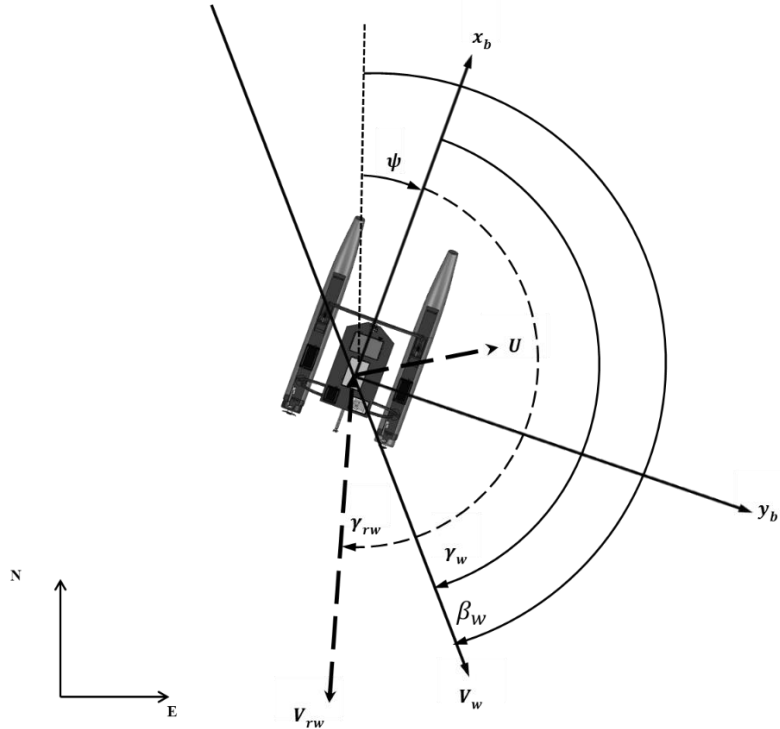

Fig. 18: Representation of vehicle heading ($\psi$), velocity ($\boldsymbol{U}$), true wind speed ($V_w$), true wind direction ($\beta_w$), true wind angle of attack ($\gamma_w$), apparent wind speed ($V_{rw}$), and apparent wind angle of attack ($\gamma_{rw}$).

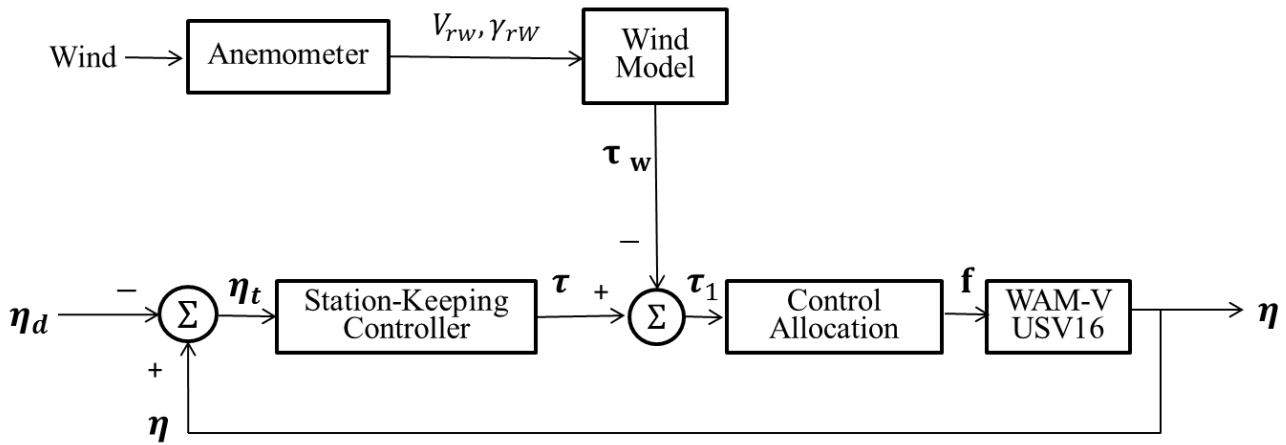

Fig. 19: Block diagram of WAM-V USV16 control system with wind feedforward.

In addition to the state feedback, the control system utilizes two additional inputs: the apparent wind speed $V_{rw}$ and direction $\gamma_{rw}$. The wind data output by the anemometer are low pass filtered, using a moving average filter with a span of 20 samples, before being input into the feedforward controller. $V_{rw}$ and $\gamma_{rw}$ are used in (31) and (32) to calculate the wind disturbance $\boldsymbol{\tau_w}$, which is then subtracted from the output of the feedback controller, leading to:

$$\boldsymbol{\tau_1} = \boldsymbol{\tau} - \boldsymbol{\tau_w}. \tag{44}$$

Here $\boldsymbol{\tau}$ is the output of the station-keeping controller, based solely on the vehicle state, obtained using either (17), (24) or (30), and $\boldsymbol{\tau_1}$ is a more accurate estimate of the output required that takes into account the effect of wind on the system $\boldsymbol{\tau_w}$. $\boldsymbol{\tau_1}$ is calculated utilizing both the anemometer output and the system state, while $\boldsymbol{\tau}$ is based solely on the system state.

By applying feedforward control theory for the purpose of improving the USV16 station-keeping performance, three additional controllers are derived. Namely, a wind feedforward PD station-keeping controller similar to the one described in Section V.A, a wind feedforward backstepping station-keeping controller similar to the one described in Section V.B and a wind feedforward sliding mode station-keeping controller similar to the one described in Section V.C. The control block for the wind model necessary for the wind feedforward controllers described is provided in Fig. 20.

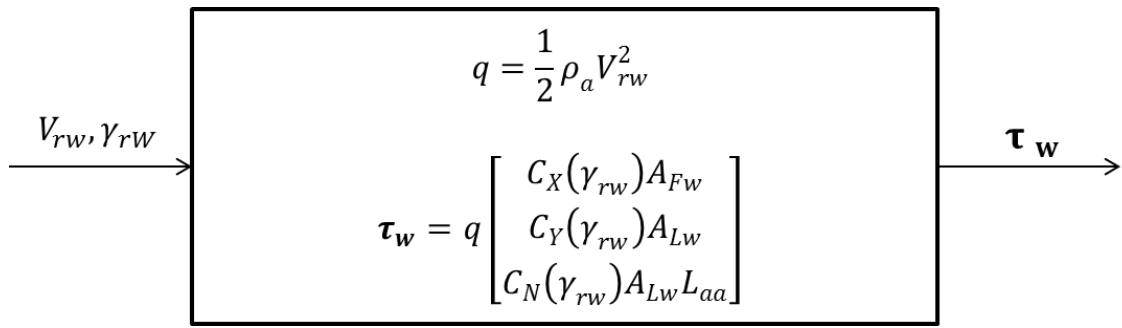

Fig. 20: Wind feedforward model.

## VI. CONTROL ALLOCATION

Azimuthing thruster configurations, such as those found on the USV16, create an overactuated system, since multiple solutions to the controller output $\boldsymbol{\tau}$ or $\boldsymbol{\tau_1}$ can be found in terms of propeller thrust and azimuth angle (Fig. 3). This is formulated as an optimization problem, and techniques such as linear programming [37], [60], quadratic programming [61], and evolutionary algorithms [62] can be used to find an optimal, or near optimal, solution. Actuator dynamics create constraints on this system, leading to a constrained nonlinear optimization problem, which is nontrivial to solve. An alternative to this method is to use a Lagrangrian multiplier technique as described in [37]. A summary of our approach is presented below.

*A.    Extended Force Representation*

For the $m$ outputs of the controller, $\boldsymbol{\tau} \in \mathbb{R}^m$, let $\mathbf{f} \in \mathbb{R}^{2k}$ be the actuator forces in the surge and sway directions at each of the $k$ actuators,

$$\mathbf{f} = \begin{bmatrix} F_{x_1} \, F_{y_1} \, ... \, F_{x_i} \, F_{y_i} \, ... \, F_{x_k} \, F_{y_k} \end{bmatrix}^T. \tag{45}$$

A transformation matrix $\boldsymbol{T} \in \mathbb{R}^{2k \times m}$ from the controller output force $\boldsymbol{\tau}$ (or $\boldsymbol{\tau_1}$) to the actuator frame force vector $\mathbf{f}$ can be defined as

$$\boldsymbol{\tau} = \boldsymbol{T}\mathbf{f}, \tag{46}$$

where $\boldsymbol{T}$ is generically defined in [63] as:

$$\mathbf{T} = \begin{bmatrix} 1 & 0 & ... & 1 & 0 \\ 0 & 1 & ... & 0 & 1 \\ -l_{y_1} & l_{x_1} & ... & -l_{y_k} & l_{x_k} \end{bmatrix}. \tag{47}$$

The constants $l_{x_i}$ and $l_{y_i}$ represent the longitudinal and lateral distances to the $i$th actuator measured with respect to the vehicle center of gravity. The propulsion system mounted on the USV16 (Fig. 2) consisted of two linear actuators and two thrusters, therefore $k = 4$. Thus, the generic equations (45) and (47) are rewritten to define $\mathbf{f}$ and $\boldsymbol{T}$ for the system used:

$$\mathbf{f} = \begin{bmatrix} F_{x_p} \, F_{y_p} \, F_{x_s} \, F_{y_s} \end{bmatrix}^T, \tag{48}$$

$$\mathbf{T} = \begin{bmatrix} 1 & 0 & 1 & 0 \\ 0 & 1 & 0 & 1 \\ -l_{y_p} & l_{x_p} & -l_{y_s} & l_{x_s} \end{bmatrix}. \tag{49}$$

In (48) and (49) the subscripts $s$ and $p$ stand for the starboard and port sides, respectively. The solution to the allocation problem now rests in finding an inverse to the rectangular transformation matrix $\boldsymbol{T}$.

*B.    Lagrangian Multiplier Solution*

A cost function $C$ is set up to minimize the force output from each actuator subject to a positive definite weight matrix $\mathbf{W} \in \mathbb{R}^{2k \times 2k}$,

$$\min_{\mathbf{f}} \{ C = \mathbf{f}^{\mathbf{T}} \mathbf{W} \mathbf{f} \}, \tag{50}$$

The optimization problem in (50) is subject to the constraint $\boldsymbol{\tau} - \mathbf{T}\mathbf{f} = \mathbf{0}$, i.e., the error between the desired control forces and the attainable control forces is minimized. The weight matrix $\mathbf{W}$ is set to skew the control forces towards the most efficient actuators. This is especially important for systems with rudders or control fins, as these actuators provide greater control authority with less power consumption.

A Lagrangrian is then set up as in [59],

$$L(\mathbf{f}, \boldsymbol{\lambda}) = \mathbf{f}^{\mathbf{T}} \mathbf{W} \mathbf{f} + \boldsymbol{\lambda}^{\mathbf{T}} (\boldsymbol{\tau} - \mathbf{T}\mathbf{f}). \tag{51}$$

Differentiating (51) with respect to $\mathbf{f}$, one can show that the solution for $\mathbf{f}$ reduces to $\mathbf{f} = \mathbf{T}_w^\dagger \boldsymbol{\tau}$, where the inverse of the weighted transformation matrix is,

$$\mathbf{T}_w^\dagger = \mathbf{W}^{-1}\mathbf{T}^{\mathrm{T}}(\mathbf{T}\mathbf{W}^{-1}\mathbf{T}^{\mathrm{T}})^{-1}. \tag{52}$$

If a vehicle has port/starboard symmetry with identical actuators, the weight matrix $\mathbf{W}$ can be taken as the identity matrix, $\mathbf{W} = \mathbf{I} \in \mathbb{R}^{2k \times 2k}$, and the inverse of the transformation matrix becomes the Moore-Penrose pseudoinverse of the transformation matrix, $\mathbf{T}_w^\dagger = \mathbf{T}^{\mathrm{T}}(\mathbf{T}\mathbf{T}^{\mathrm{T}})^{-1}$.

Once the component force vector $\mathbf{f}$ is found, a four-quadrant *arctan* function can be applied to find the desired azimuth angles $\delta_p = \tan^{-1}(F_{y_p}/F_{x_p})$ and $\delta_s = \tan^{-1}(F_{y_s}/F_{x_s})$, and to calculate the magnitude of the thrust at each propeller $T_p = \sqrt{F_{x_p}^2 + F_{y_p}^2}$ and $T_s = \sqrt{F_{x_s}^2 + F_{y_s}^2}$. The block for the control allocation described is provided in Fig. 21.

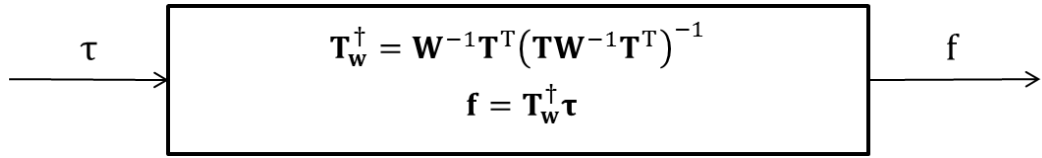

$\boldsymbol{\tau}$ $\qquad$ $\mathbf{T}_w^\dagger = \mathbf{W}^{-1}\mathbf{T}^{\mathrm{T}}(\mathbf{T}\mathbf{W}^{-1}\mathbf{T}^{\mathrm{T}})^{-1}$
$\mathbf{f} = \mathbf{T}_w^\dagger \boldsymbol{\tau}$ $\qquad$ $\mathbf{f}$

Fig. 21: Control allocation using the extended thrust representation to convert from desired forces $\boldsymbol{\tau}$ to an extended thrust representation $\mathbf{f}$

Owing to physical limitations on the travel of the linear actuators, the azimuth range of each propeller is from -45° to +45°. However, a 180° offset from a value in this range is also attainable by reversing the propeller. A logic scheme is implemented on top of the control allocation that sets the thrust to zero if the allocation scheme requests an unachievable angle, and reverses it if an angle from -135° to 135° is required. This scheme is illustrated in Fig. 22. Careful tuning of controller parameters is necessary to ensure that these constraints are not violated. The approach produces a computationally efficient answer to the overallocation optimization problem, which is possible to implement on the USV's embedded processor.

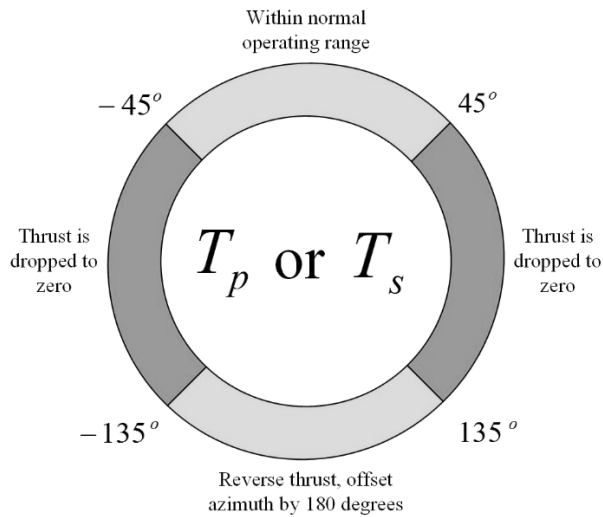

Within normal operating range

$-45^o$ $\qquad$ $45^o$

Thrust is dropped to zero $\qquad$ $T_p$ or $T_s$ $\qquad$ Thrust is dropped to zero

$-135^o$ $\qquad$ $135^o$

Reverse thrust, offset azimuth by 180 degrees

Fig. 22: Control allocation logic reverses thrust $T_p = -T_p$ when $135 < |\delta_p| < 180$ and $T_s = -T_s$ when $135 < |\delta_s| < 180$. Thrust is set to zero, $T_p = 0$ when $45 < |\delta_p| < 135$ and $T_s = 0$ when $45 < |\delta_s| < 135$.

Due to the fact that the time responses of the thrusters and linear actuators aren't precisely modeled within the allocation scheme, the resultant forces and angles commanded are low-pass filtered with a user-set cutoff frequency to maintain a feasible response from the propulsion system. The low-pass filters used here are simple first order, infinite impulse response filters with a single time constant that is used to set the cut-off frequency. The time constant is set to conservatively match the actuator dynamics. Two separate low-pass filters were used to filter the control allocation output for the azimuth as well as thrust for each motor. The time constant for the thrust filter was found to be about an order of magnitude greater than that of the azimuth filter.

VII.    STATION-KEEPING EXPERIMENTS

A series of sea trials were performed in calm water sections of the U.S. Intracoastal Waterway in Dania Beach, FL to test the performance of the station-keeping controllers described in Section V and the control allocation scheme in Section VI. In order for the vehicle to operate effectively, it cannot be used in sea states greater than sea state 1, wind speeds greater than 15 knots, wave heights greater than 0.2 meters and environments with heavy currents. To minimize the effect of waves on the vehicle, no tests were performed on the open ocean. Wind and current were therefore the two major causes of environmental disturbance. The effect of waves on the vehicle was neglected. While the wind disturbance on the vehicle could be estimated, and accounted for, as explained in Section V.D, the USV16 currently lacks of an appropriate sensor to measure water currents.

To provide an estimate of the environmental disturbance caused by tidal currents to which the USV was exposed to during sea trials, the tidal level, recorded at the nearby (~4 km) Turning Basin of Port Everglades, Fort Lauderdale, FL, over the course of the days when the data were collected at each location, is shown in Fig. 23. The effect of the tidal current disturbance on the vehicle was not estimated or measured as part of this research, however, based on the results, it affected the performance of the controller during station-keeping trials.

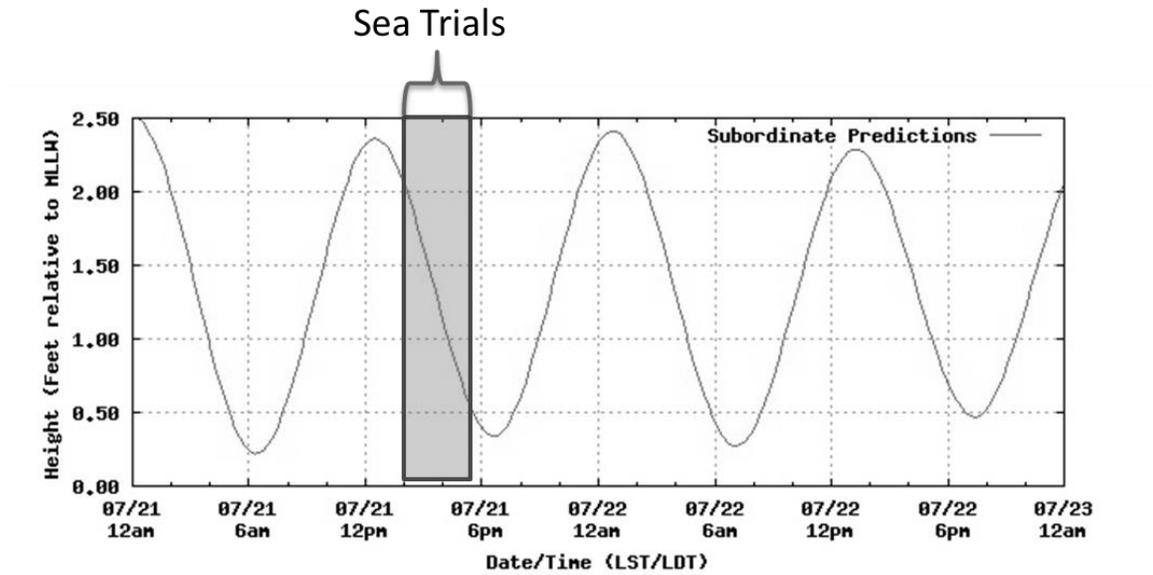

Fig. 23: Tide during sea trials recorded at the Port Everglades Turning Basin, Fort Lauderdale, FL. As indicated, the measurements were conducted during an outgoing (ebb) tide. Tide height is referenced to the mean lower low water level (MLLW), which is the height of the lowest tide recorded on the date of the measurement. The acronym LST/LDT indicates the local standard time adjusted to daylight savings time.

Two different locations were chosen to perform the experiments; each controller described in Section V was therefore tested at each of these. A small map illustrating the desired state of the vehicle at each location is shown in Fig. 24. The locations were selected to expose the vehicle to minimum (Location 1) and maximum (Location 2) local environmental disturbances. The desired heading of the vehicle was also chosen to reproduce the most friendly (Location 1) and the harshest (Location 2) scenarios.

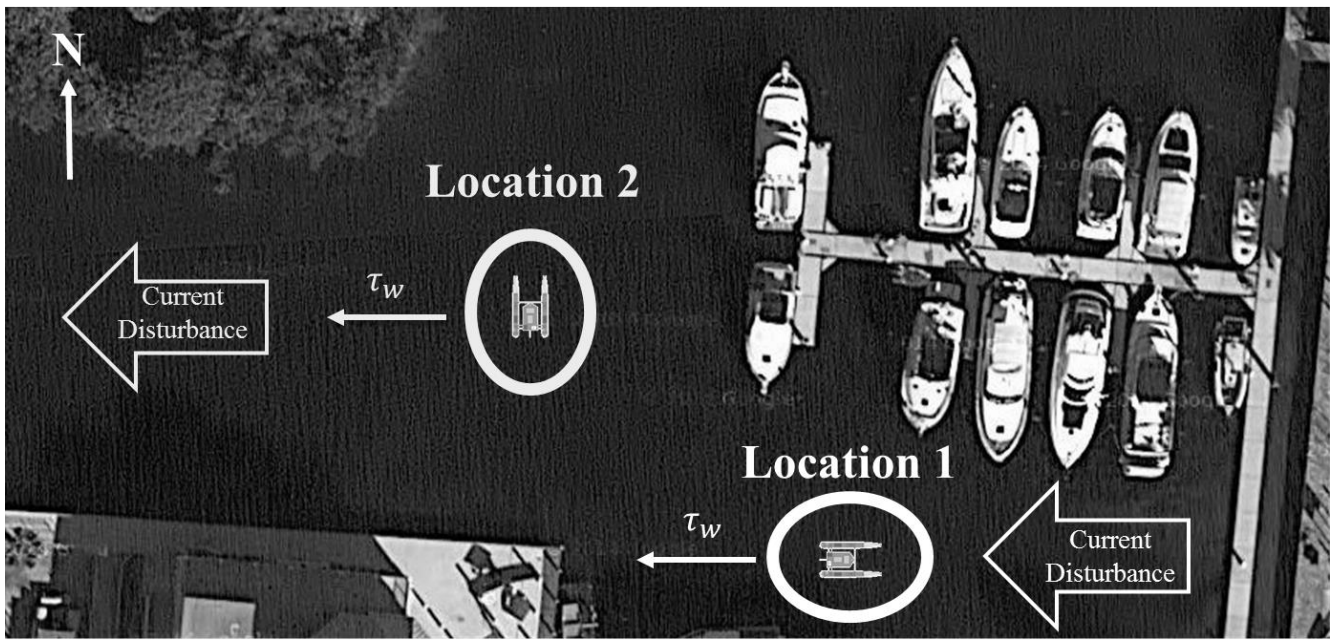

Fig. 24: Test locations. Location 1 was more sheltered in comparison to Location 2. Location 2 exposed the vehicle to greater current disturbance due to the channel created between the northern and western inlets.

The USV16 is designed to navigate primarily in the surge direction; therefore this type of motion is considered the simplest to achieve. At Location 1, the vehicle's desired heading is in the opposite direction of the disturbance, therefore the controller will output a high force in the surge direction, $T_x$ in (12). This maneuver can be easily achieved and is consistent with a "friendly" scenario. As the azimuth angles available to the propulsion system are limited (Section VI.B), the most difficult configuration to achieve will be one that requires a large sway force $T_y$, while simultaneously commanding a torque $M_z$. At Location 2 the vehicle's desired heading is 90º from the disturbance direction. Therefore, the controller will output a high force in the sway direction, $T_y$ in (12), to counteract the disturbance, creating the harshest scenario.

The apparent wind speed $V_{rw}$ and apparent wind direction $\gamma_{rw}$ was recorded directly from the vehicle during the experiments to quantify the sensed disturbance. Mean and standard deviation of $V_{rw}$ and $\gamma_{rw}$ at Location 1 and Location 2 over the time at which each controller was being tested is shown in TABLE 4 and TABLE 5 respectively. Note that all the mean values of $V_{rw}$ have similar magnitudes at both locations. However, the mean values of $\gamma_{rw}$ average around 230º at Location 2 and 180º at Location 1, (meaning that the wind drag on the USV had a substantial transverse component at Location 2). For these reasons, and due to the tidal current direction, Location 2 presented a harsher environment for station-keeping.

TABLE 4

MEAN AND STANDARD DEVIATION OF APPARENT WIND SPEED AND APPARENT WIND, AND WIND TURBULENCE DIRECTION AT LOCATION 1 FOR PD, BACKSTEPPING AND SLIDING MODE STATION-KEEPING CONTROLLERS WITH AND WITHOUT WIND FEEDFORWARD CONTROL, THE DURATION IS 700 S.

| Controller: | PD | PD with feedforward | Backstepping | Backstepping with feedforward | Sliding Mode | Sliding Mode with feedforward |
|---|---|---|---|---|---|---|
| Mean Apparent Wind Speed (m/s): | 2.42 | 2.36 | 2.46 | 2.10 | 2.17 | 2.35 |
| Standard Deviation of Apparent Wind Speed (m/s) | 0.68 | 0.70 | 0.62 | 0.64 | 0.66 | 0.76 |
| Mean Apparent Wind Direction (deg) | 180.4 | 180.5 | 171.3 | 176.5 | 179.5 | 177.6 |
| Standard Deviation of Apparent Wind Direction (deg): | 15.0 | 14.0 | 14.1 | 18.4 | 13.8 | 14.2 |
| Wind Turbulence Intensity $s \equiv TKE/\bar{V}_{rw}$ (%) | 14.1 | 14.8 | 12.6 | 15.2 | 15.3 | 16.1 |

TABLE 5

MEAN AND STANDARD DEVIATION OF APPARENT WIND SPEED AND APPARENT WIND DIRECTION, AND WIND TURBULENCE AT LOCATION 2 FOR PD, BACKSTEPPING AND SLIDING MODE STATION-KEEPING CONTROLLERS WITH AND WITHOUT WIND FEEDFORWARD CONTROL, THE DURATION IS 700S.

| Controller: | PD | PD with feedforward | Backstepping | Backstepping with feedforward | Sliding Mode | Sliding Mode with feedforward |
|---|---|---|---|---|---|---|
| Mean Apparent Wind Speed (m/s): | 2.86 | 2.47 | 2.43 | 2.59 | 2.14 | 1.98 |
| Standard Deviation of Apparent Wind Speed (m/s) | 0.64 | 0.57 | 0.59 | 0.54 | 0.73 | 0.80 |
| Mean Apparent Wind Direction (deg) | 222.4 | 225.6 | 229.6 | 234.55 | 215.9 | 216.0 |
| Standard Deviation of Apparent Wind Direction (deg): | 17.25 | 18.0 | 20.4 | 18.8 | 28.89 | 40.44 |
| Wind Turbulence Intensity $s \equiv TKE/\bar{V}_{rw}$ (%) | 11.1 | 11.5 | 12.1 | 10.4 | 17.1 | 20.2 |

The performance of each controller was tested with and without the wind feedforward feature. All experiments were initialized by bringing the vehicle to its desired state manually using a remote controller outfitted on the GNC hardware. The system was then commanded to engage in autonomous mode and maintain its state for 700 seconds. This procedure allowed the controller to act on the vehicle at steady-state conditions with zero initial error. Previous sea trials showed that, if the station-keeping command was given with an initial error in heading and position, all controllers were able to drive the system to steady-state rapidly. Heading and position were recorded throughout each run. To evaluate the performance of each controller, the errors in both heading and position were plotted. To evaluate the effectiveness of the wind feedforward feature, the state error was plotted for the same station-keeping controller with and without the wind feedforward control portion.

*A.     Station-keeping Trials at Location 1: Vehicle Friendly Scenario*

As previously mentioned, the PD controller was implemented to provide a baseline for the development and testing of the other nonlinear robust controllers.  Position and heading error for the PD station-keeping controller, with and without the wind feedforward control feature, operating at Location 1, are shown in Fig. 25.

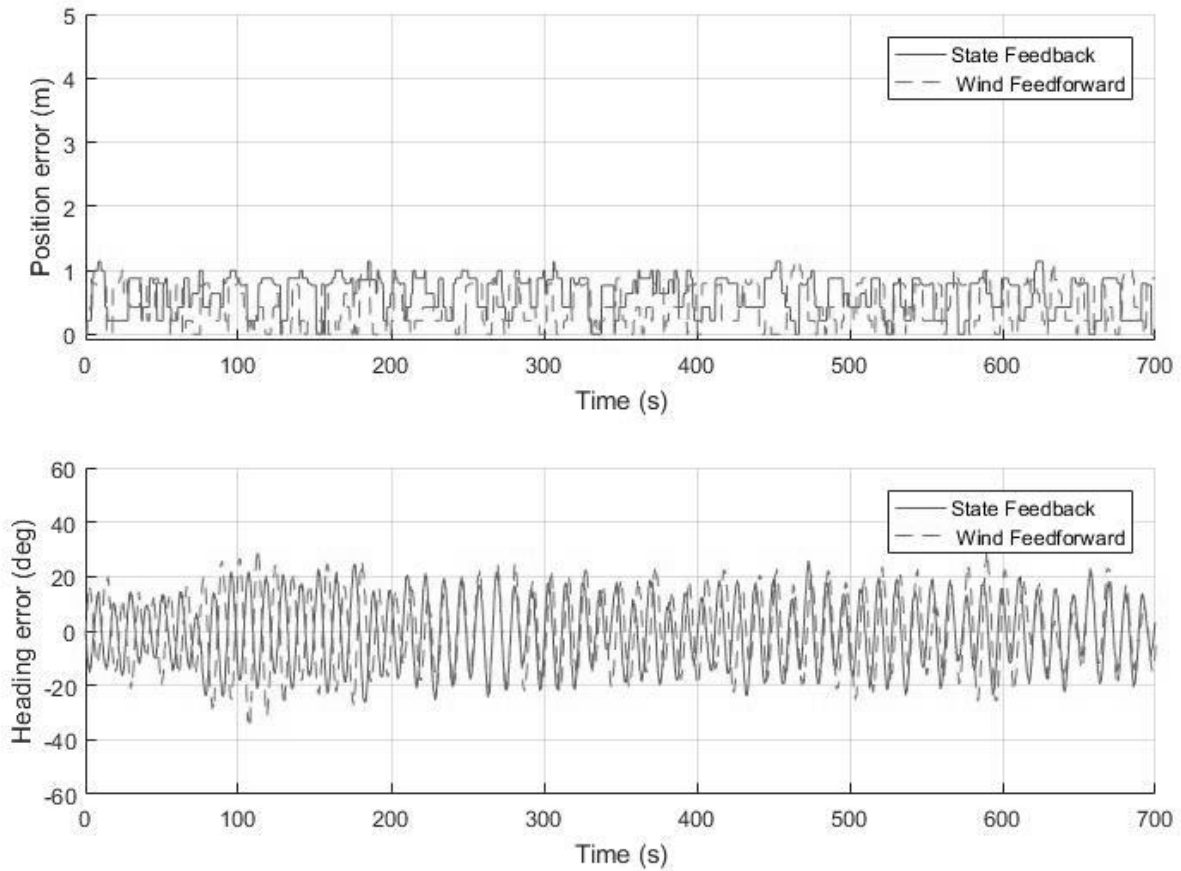

Fig. 25: Position and heading errors for PD station-keeping controller with/without wind feedforward control at Location 1. As the resolution of the GPS sensor is about 1 m, the position measurements exhibit a step-like appearance.

It can be seen that, while the USV position error is maintained below 1 meter most of the time, the heading error steadily fluctuates between -20º and +20º. It is important to note that the PD station-keeping controller always demonstrated a similar limited functionality. The PD controller's capability to station-keep the vehicle was only achievable by setting a very high heading gain. This resulted in a highly oscillating heading error, caused by the large control effort. The steady oscillation of the vehicle's heading enabled it to hold position, maintaining the position error always at or below the GPS sensor accuracy of 1 meter. Gain tuning techniques were not sufficient to reduce the heading error; minimal improvements in the heading performance caused a drastic increase of position error, resulting in an inability of the vehicle to hold position. Despite significant manual tuning, adding the wind feedforward portion to the PD station-keeping controller simply resulted in larger oscillations of the vehicle's heading, while the position error was not affected.

The first robust controller implemented on the USV16 for the purpose of station-keeping was the backstepping controller described in Section V.B. The results for the backstepping station-keeping controller, with and without the wind feedforward control portion, operating at Location 1 are shown in Fig. 26.

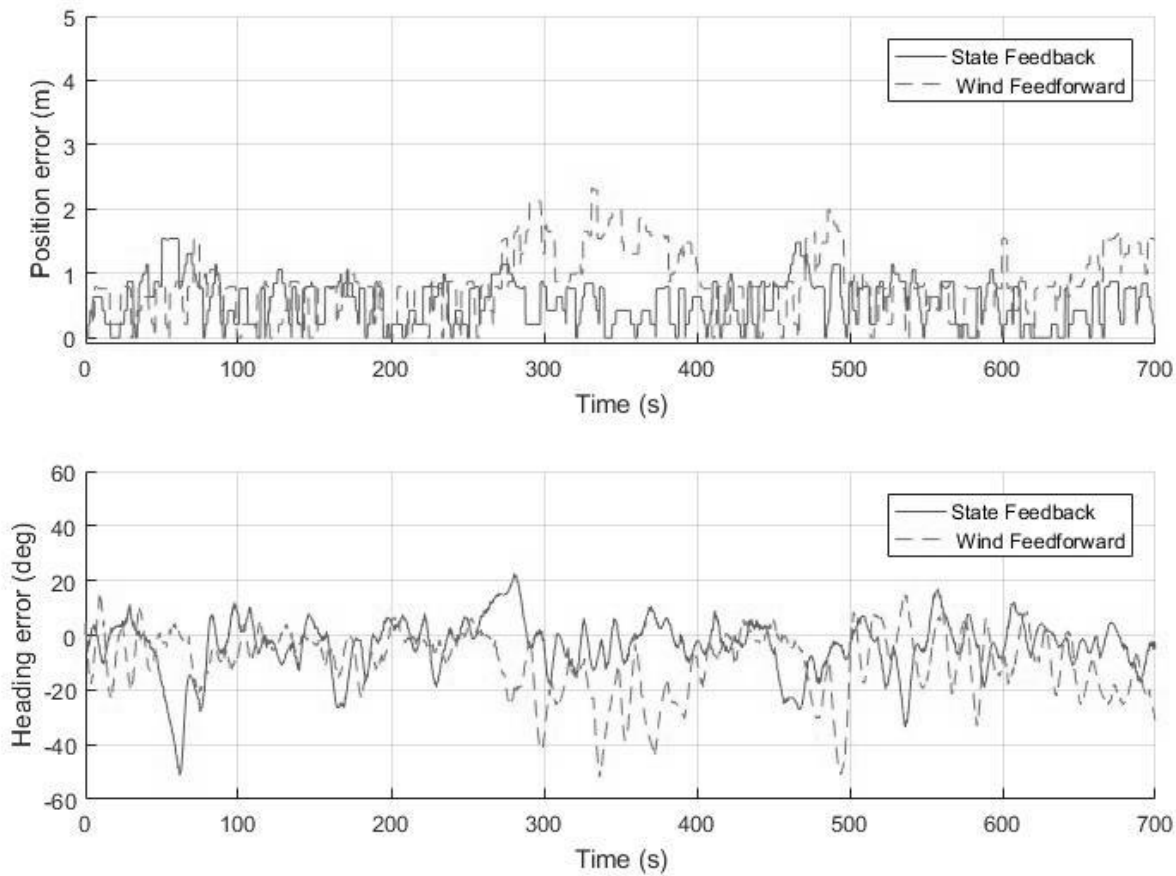

Fig. 26: Position and heading errors for backstepping station-keeping controller with/without wind feedforward control at Location 1. As the resolution of the GPS sensor is about 1 m, the position measurements exhibit a step-like appearance.

Applying backstepping robust techniques for the purpose of station-keeping enables the USV to maintain position and heading over an extended period of time. As a result, the heading error is minimized while still maintaining a small (~1m) position error. Differently from the PD controller, the capability of the vehicle to hold position is no longer dependent on the need to maintain high and steady oscillations of the heading parameter. The rate at which the heading error varies is also heavily reduced, when utilizing the backstepping controller. The addition of wind feedforward control did not cause any relevant variation in the performance of the backstepping station-keeping controller.

The second robust controller implemented on the USV16 for the purpose of station-keeping was the sliding mode controller described in section V.C. In addition to the sliding mode theory, an integral term was added in (28) and (29) to remove any steady-state error present when testing the backstepping station-keeping controller. The results for the sliding mode station-keeping controller, with and without the wind feedforward control portion, operating at Location 1 are shown in Fig. 27.

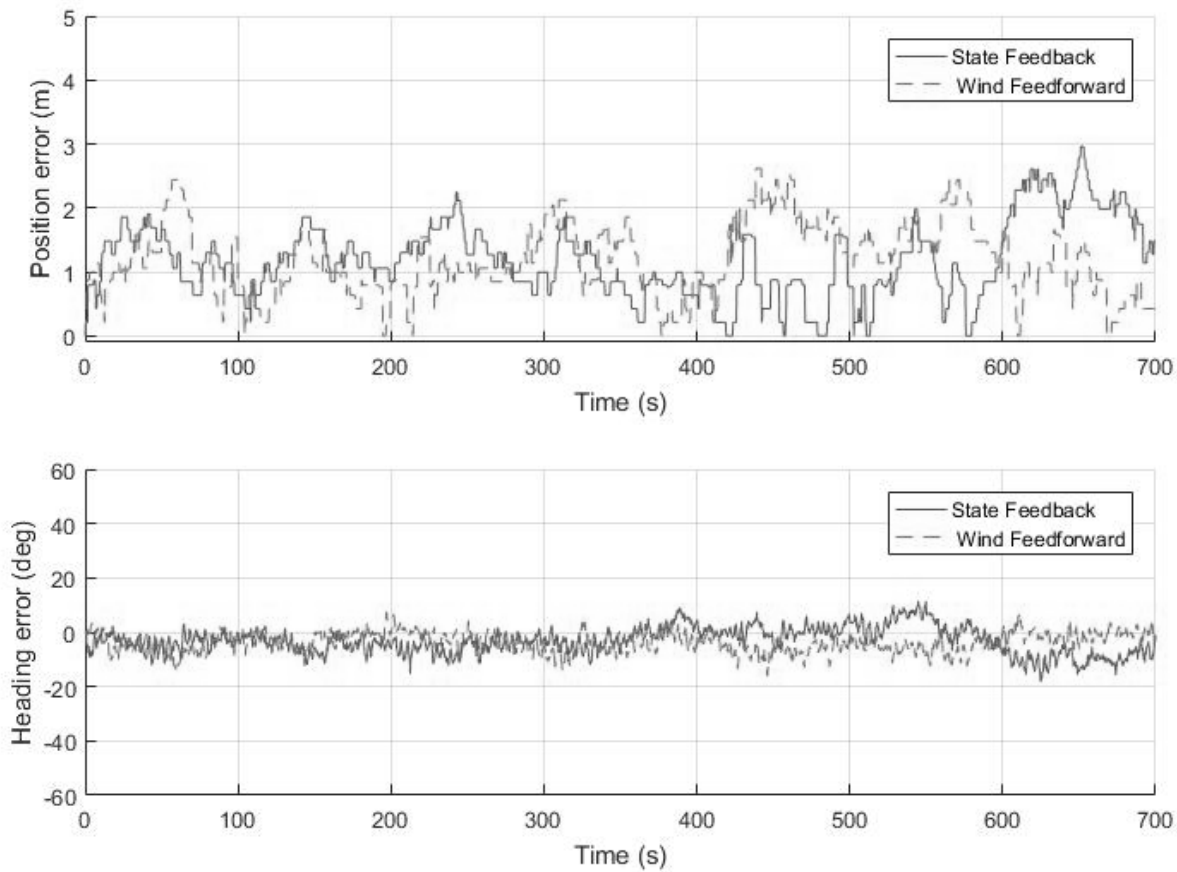

Fig. 27: Position and heading errors for sliding mode station-keeping controller with/without wind feedforward control at Location 1. As the resolution of the GPS sensor is about 1 m, the position measurements exhibit a step-like appearance.

The siding mode controller outperforms both the PD and the backstepping controllers in heading performance, without a significant increase in position error. Chattering behavior is evident in both data plots in Fig. 27. Appropriate tuning of the sliding mode parameters was essential to maximize the performance of the sliding mode controller. More specifically, the boundary layer thickness, $E$ in (30), had to be precisely defined, since excessively reducing its values caused constant saturation of the control output resulting in large errors, and excessively increasing it caused uncontrolled chattering that deviated the system from its desired state. The bound around the uncertainties, $R$ in (30) also had to be tuned accordingly, based on the controller performance. Extensive testing was carried out to manually refine these parameters, until the final values were identified. The improvement in performance was then evident, since the error was contained and slowly varied within the boundary layer thickness throughout the experiments. The performance of the sliding mode station-keeping controller could not be improved any further by applying wind feedforward control.

The performance of the station-keeping controllers, is compared in TABLE 6.

TABLE 6

MEAN AND STANDARD DEVIATION OF POSITION AND HEADING ERROR AT LOCATION 1 FOR PD, BACKSTEPPING
AND SLIDING MODE STATION-KEEPING CONTROLLERS WITH/WITHOUT WIND FEEDFORWARD CONTROL.

| Controller: | PD | PD with feedforward | Backstepping | Backstepping with feedforward | Sliding Mode | Sliding Mode with feedforward |
|---|---|---|---|---|---|---|
| Mean Position Error (m): | 0.61 | 0.43 | 0.55 | 0.84 | 1.18 | 1.18 |
| Standard Deviation of Position Error (m) | 0.30 | 0.33 | 0.36 | 0.51 | 0.59 | 0.56 |
| Mean Heading Error (deg) | 10.91 | 11.86 | 7.66 | 11.52 | 4.69 | 4.28 |
| Standard Deviation of Heading Error (deg): | 6.08 | 6.87 | 7.58 | 10.43 | 3.37 | 3.07 |

*B.      Station-keeping Trials at Location 2: Vehicle Unfriendly Scenario*

At Location 2 the vehicle is exposed to tranverse current and wind disturbances, which the propulsion system can't counteract while maintaining the same USV heading $\psi$. This forces the vehicle to momentarily deviate from steady-state, leading to larger errors. Specifically, as position error in the y direction increases, the position error in the x direction and the heading error are maintained low. This scenario results in a controller output, $\boldsymbol{\tau}$ in (12), with very small surge force $T_x$ and moment $M_z$, but a very large sway force $T_y$. The required propulsion system configuration, $\mathbf{f}$ in (48), then becomes very difficult to achieve due to the constraints in thrust and azimuthing angles. Position and heading error for the PD station-keeping controller, with and without the wind feedforward control portion, operating at Location 2, are shown in Fig. 28.

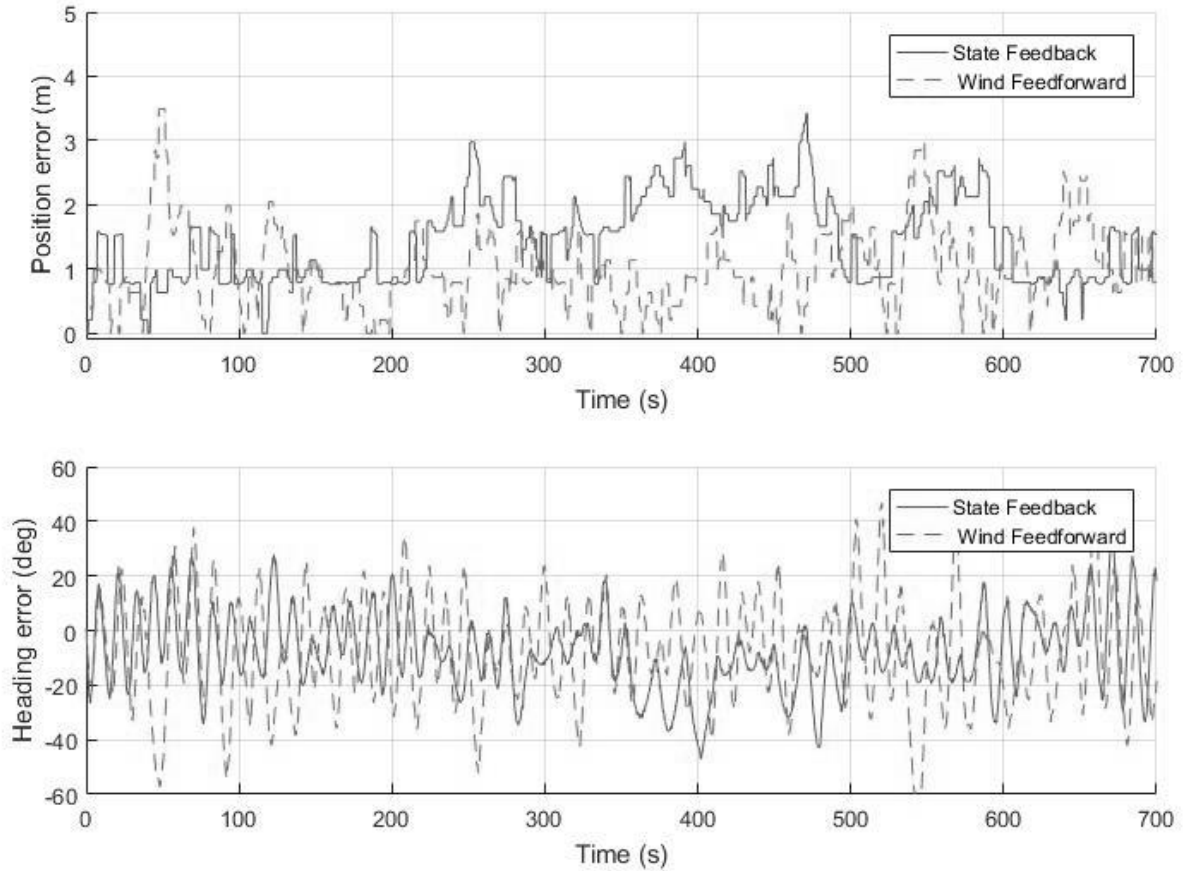

Fig. 28: Position and heading errors for PD station-keeping controller with/without wind feedforward control at Location 2.

The results for the PD controller at Location 2 are consistent with what was observed at Location1: larger and consistent heading oscillations enable the vehicle to maintain position. As expected, the errors in both heading and position are larger than at Location 1 and the wind feedforward feature does not improve the vehicle's station-keeping capability.

It was expected that the backstepping controller would improve the station-keeping performance in maintaining the vehicle's heading, while still presenting a larger position error due to direction of the disturbance. The results for the backstepping station-keeping controller, with and without the wind feedforward control portion, operating at Location 2 are shown in Fig. 29.

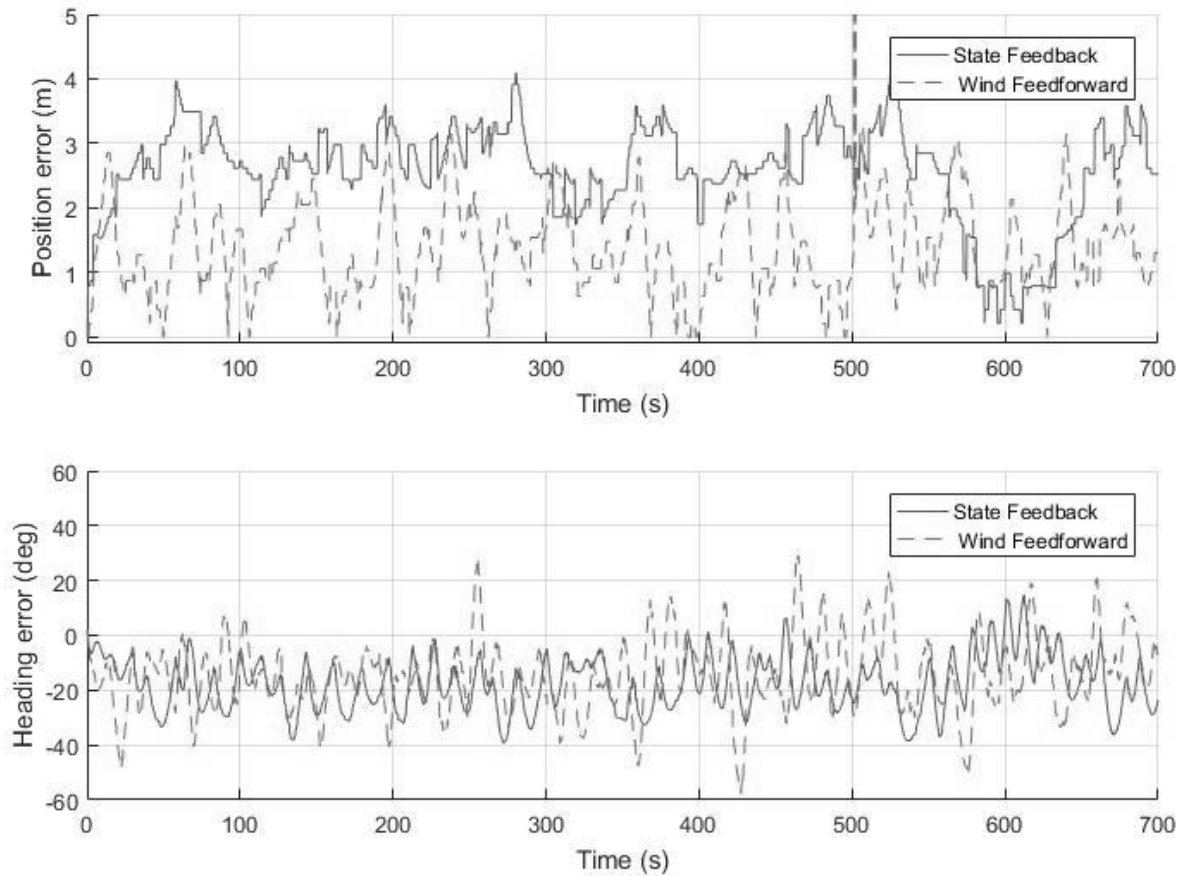

Fig. 29: Position and heading errors for backstepping station-keeping controller with/without wind feedforward control at Location 2.

The backstepping performance was heavily compromised at Location 2. As expected, the error in position increased dramatically as compared to Location 1. The heading error also seems to fluctuate similarly to the PD controller, however the heading oscillation observable when testing the backstepping controller are mostly maintained between 0º and -40º. This proves that the vehicle is commanded to momentarily vary its heading in order to fix its position error, but once the heading is brought back to its desired value, the USV quickly loses its desired position again. In other words, it is shown that, as the position error quickly increases, the vehicle is forced to sacrifice its heading to recover, resulting in continuous variations of the heading error. Applying wind feedforward control substantially helps the station-keeping performance of the backstepping controller. Although peaks in heading error can be spotted when applying wind feedforward control, the vehicle is able to always correct the position error. Instead, when wind feedforward control is not applied, a steady-state position error of 3 m can be identified in Fig. 29, meaning that the backstepping controller alone is not able to correct the vehicle position.

It was expected that the sliding mode controller would improve the USV station-keeping performance in both heading and position. Also, similarly as for the backstepping, it was predicted that wind feedforward control would benefit the sliding mode controller. The results for the sliding mode station-keeping controller, with and without the wind feedforward control, operating at Location 2 are shown in Fig. 30.

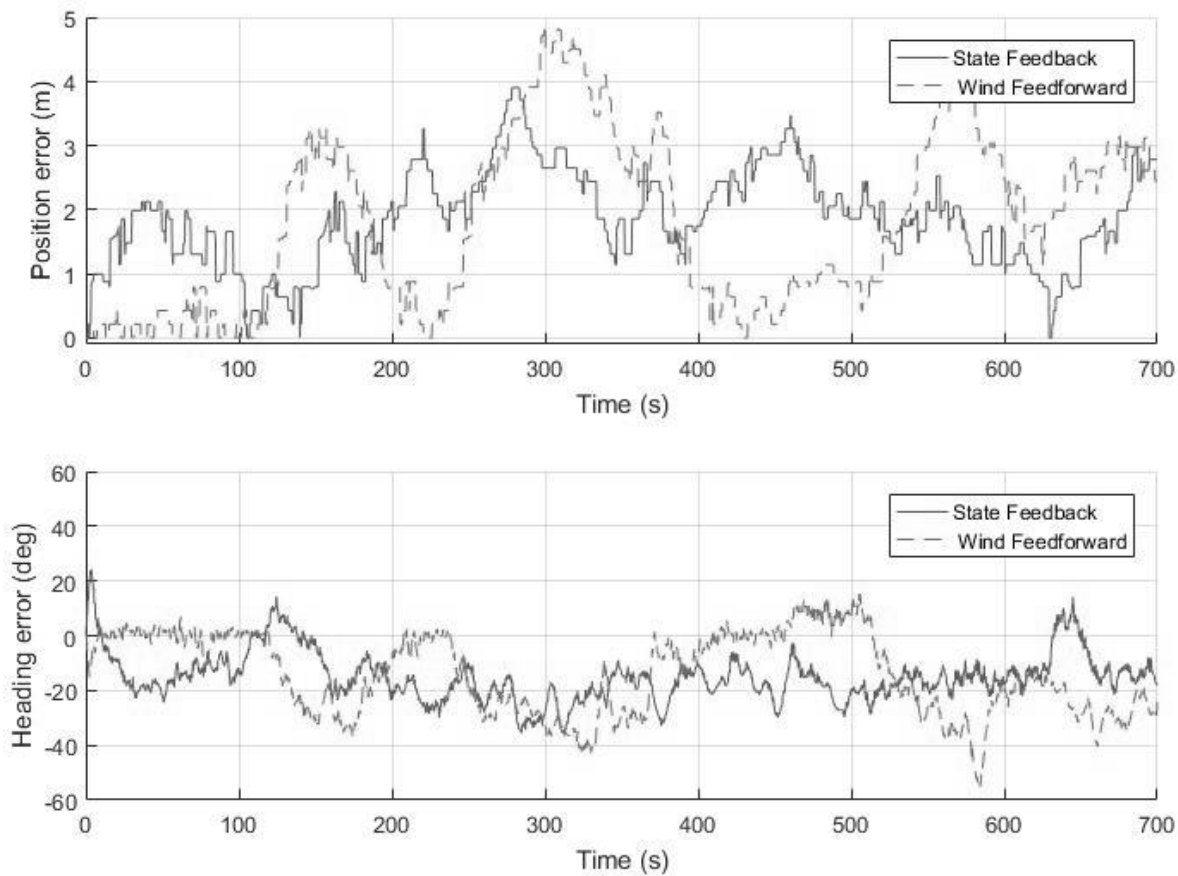

Fig. 30: Position and heading errors for sliding mode station-keeping controller with/without wind feedforward control at Location 2.

At Location 2, the sliding mode controller also outperforms both the PD and backstepping controllers. Differently from the other controllers, the errors in both position and heading increase very slowly over time and they are always brought back to zero. However, while sliding mode control theory forces the system to reach and stay within the boundary layer, it does not guarantee a quick response to an increase in error, which causes the system exit the boundary layer [64]. As expected, when applying wind feedforward control, the errors are further minimized.

Mean and standard deviation of position and heading error at Location 2 are presented for each station-keeping controller in TABLE 7.

TABLE 7

MEAN AND STANDARD DEVIATION OF POSITION AND HEADING ERROR AT LOCATION 2 FOR PD, BACKSTEPPING
AND SLIDING MODE STATION-KEEPING CONTROLLERS WITH/WITHOUT WIND FEEDFORWARD CONTROL.

| Controller: | PD | PD with feedforward | Backstepping | Backstepping with feedforward | Sliding Mode | Sliding Mode with feedforward |
|---|---|---|---|---|---|---|
| Mean Position Error (m): | 1.45 | 1.01 | 2.57 | 1.49 | 1.85 | 1.80 |
| Standard Deviation of Position Error (m) | 0.66 | 0.62 | 0.77 | 0.77 | 0.75 | 1.35 |
| Mean Heading Error (deg) | 12.9 | 16.9 | 17.64 | 17.00 | 15.96 | 16.03 |
| Standard Deviation of Heading Error (deg): | 9.21 | 12.36 | 9.07 | 10.37 | 7.08 | 12.85 |

## VIII.    SUMMARY AND CONCLUDING REMARKS

The results from experimental station keeping tests of a USV using either pure feedback control, or feedback control coupled with wind feedforward control, are presented. The tests were conducted to evaluate the performance of proportional derivative (PD), backstepping and sliding mode feedback controllers, with and without wind feedforward control, while the USV was exposed to wind and current disturbances.

The nonlinear PD controller is used as a basis for comparison. The backstepping controller derived in [22] for large mono-hull vessels is adapted for a small, twin-hulled USV. A sliding mode controller, similar to the one described in [48], is then developed to reduce steady-state errors observed with the PD and backstepping controllers. Lastly, to account for the quickly varying wind disturbances experienced during field trials, a novel wind feedforward controller was derived and implemented to improve the performance of the other controllers. A comparative analysis of the steady-state performance of the station-keeping controllers developed here was performed at two different locations, exposing the vehicle to similar disturbances, but which acted across different vehicles axes. The major result of this effort is the development of a control approach that enables the USV to simultaneously maintain both heading and position utilizing only two transom-mounted thrusters.

The results show that the sliding mode feedback controller performed best overall and that the addition of wind feedforward control did not significantly improve its effectiveness. However, wind feedforward control did substantially improve the performance of the proportional derivative and backstepping controllers when the mean wind direction was perpendicular to the longitudinal axis of the USV. When wind and current disturbances are aligned with the longitudinal (surge) axis of the vehicle, the wind feedforward control does not significantly improve the performance of the station keeping controllers.

The approach to station keeping control taken in this research led to five major conclusions:

i.    For twin hull USVs operating in environmental conditions similar to the ones in which the vehicle was tested, two azimuthing thrusters are sufficient to enable the USV's capability of station keeping heading and position simultaneously.

ii. Using nonlinear PD or backstepping feedback control alone does not provide sufficient control effort to station-keep both the heading and position, simultaneously.

iii. A compromise exists between station keeping heading vs position. In addition to gain, or parameter, tuning techniques to refine this trade-off based on the application, the selection of an appropriate station keeping controller (such as the sliding mode controller) can minimize the errors that arise from this compromise.

iv. The spatial variation of wind speed and direction are generally determined by the dominant length scales of the turbulence in the wind. Using representative wind data, the dominant turbulent length scales are shown to be much larger than the size of the USV used here, so that a single anemometer is deemed sufficient for wind feedforward control.

v. For a light-weight USV using nonlinear PD or backstepping control, the addition of wind feedforward control works better against transverse disturbances, than it does for longitudinal ones. This suggests that a feedforward control strategy that weights the feedforward term $\tau_w$ in (44) based on the dominant wind direction may be beneficial.

ACKNOWLEDGEMENTS

I. R. Bertaska gratefully acknowledges the support of the Link Foundation Ocean Engineering and Instrumentation Ph.D. Fellowship Program. All authors would like to thank the Link Foundation for partial support of this effort.

REFERENCES

[1] J. E. Manley, "Unmanned surface vehicles, 15 years of development.," in *MTS/IEEE OCEANS 2008*, Quebec City, QC Canada, 2008.

[2] I. R. Bertaska, J. Alvarez and et.al., "Experimental Evaluation of Approach Behavior for Autonomous Surface Vehicles," in *Proc. ASME Dynamic Systems Control Conference*, 2013.

[3] I. R. Bertaska, B. Shah, K. D. von Ellenrieder, P. Svec, W. Klinger, A. J. Sinisterra, M. R. Dhanak and S. K. Gupta, "Experimental evaluation of automatically-generated behaviors for USV operations," *Ocean Engineering*, vol. 106, pp. 496-514, 2015.

[4] C. Kitts, P. Mahacek, T. Adamek, K. Rasal, V. Howard, S. Li, A. Badaoui, W. Kirkwood, G. Wheat and S. Hulme, "Field operation of a robotic small waterplane area twin hull boat for shallow-water bathymetric characterization," *Journal of Field Robotics*, vol. 29, no. 6, pp. 924-938, 2012.

[5] R. R. Murphy, E. Steimle, M. Hall, M. Lindemuth, D. Trejo, S. Hurlebaus, Z. Medina-Cetina and D. Slocum, "Robot-Assisted Bridge Inspection," *Journal of Intelligent & Robotic Systems*, vol. 64, pp. 77-95, 2011.

[6] E. I. Sarda, M. D. Dhanak and K. D. von Ellenrieder, "Concept for a USV-based autonomous launch and recovery system," in *ASNE Launch & Recovery*, Linthicum, 2014.

[7] J. Busquets, J. V. Busquets, D. Tudela, F. Perez, J. Busquets-Carbonell, A. Barbera, C. Rodriguez, A. J. Garcia and J. Gilabert, "Low-cost AUV based on Arduino open source microcontroller board for oceanographic research applications in a collaborative long term deployment missions and suitable for combining with an USV as autonomous automatic recharging platform," in *IEEE/OES Autonomous Underwater Vehicles (AUV)*, 2012.

[8] G. Casalino, A. Turetta and E. Simetti, "A three-layered architecture for real time path planning and obstacle avoidance for surveillance USVs operating in harbour fields," in *MTS/IEEE Oceans*, Bremen, Germany, 2009.

[9] W. Klinger, I. R. Bertaska, J. Alvarez and K. von Ellenrieder, "Controller design challenges for waterjet propelled

unmanned surface vehicles with uncertain drag and mass properties," in *MTS/IEEE Oceans*, San Diego, CA USA, 2013.

[10] D. Pearson, P.-C. An, P. P. Beaujean, K. D. von Ellenrieder and M. D. Dhanak, "High level fuzzy logic guidance system for an unmanned surface vehicle (USV) tasked to perform autonomous lanuch and recovery (ALR) of an autonomous underwater vehicle.," in *IEEE/OES Autonomous Underwater Vehicles*, Oxford, MS, 2014.

[11] M. I. Miranda, P.-P. Beaujean, E. An and M. Dhanak, "Homing an Unmanned Underwater Vehicle Equipped with DUSBL to an Unmanned Surface Platform: A feasibility Study," in *MTS/IEEE Oceans*, San Diego, 2013.

[12] T. Huntsberger, H. Aghazarian, A. Howard and D. C. Trotz, "Stereo vision–based navigation for autonomous surface vessels," *Journal of Field Robotics,* vol. 28, no. 1, pp. 3-18, 2011.

[13] Y. Kuwata, M. T. Wolf, D. Zarzhitsky and T. L. Huntsberger, "Safe maritime autonomous navigation with COLREGS, using velocity obstacles," *IEEE Journal of Oceanic Engineering,* vol. 39, no. 1, pp. 110-119, 2014.

[14] A. Lebbad and C. Nataraj, "A Bayesian algorithm for vision based navigation of autonomous surface vehicles," in *2015 IEEE 7th International Conference on Cybernetics and Intelligent Systems (CIS) and IEEE Conference on Robotics, Automation and Mechatronics (RAM)*, Siem Reap, Cambodia, 2015.

[15] E. I. Sarda and M. D. Dhanak, "Unmanned Recovery of an AUV from a Surface Platform," in *MTS/IEEE Oceans*, San Diego, 2013.

[16] D. Schlipf, L. Y. Pao and W. C. Po, "Comparison of feedforward and model predictive control of wind turbines using LIDAR," in *IEEE 51st Annual Conference on Decision and Control (CDC)*, 2012.

[17] F. Kitamura, H. Sato, K. Shimada and T. Mikami, "Estimation of wind force acting on huge floating ocean structures," in *MTS/IEEE Oceans*, 1997.

[18] E. I. Sarda, H. Qu, I. R. Bertaska and K. D. von Ellenrieder, "Development of a USV Station-Keeping Controller," in *MTS/IEEE Oceans*, Genova, 2015.

[19] H. Qu, E. I. Sarda, I. R. Bertaska and K. D. von Ellenrieder, "Wind Feedforward Control of a USV," in *MTS/IEEE Oceans*, Genova, 2015.

[20] J. G. Marquardt, J. Alvarez and K. D. von Ellenrieder, "Characterization and System Identification of an Unmanned Amphibious Tracked Vehicle," *IEEE J. Oceanic Engineering,* vol. 39, no. 4, pp. 641-661, 2014.

[21] Y. Liao, Y. Pang and L. Wan, "Combined speed and yaw control of underactuated unmanned surface vehicles," *2nd International Asia Conference on Informatics in Control, Automation and Robotics (CAR),* vol. 1, pp. 157-161, 2010.

[22] T. I. Fossen and J. P. Strand, "Tutorial on nonlinear backstepping: applications to ship control," *Modeling, identification and control,* vol. 20.2, pp. 83-134, 1999.

[23] Ashrafiuon, Hashem and et al., "Sliding-mode tracking control of surface vessels," *Journal of Field Robotics,* vol. 30.3, pp. 371-398, 2013.

[24] Sonnenburg, R. Christian and C. A. Woolsey, "Modeling, identification, and control of an unmanned surface vehicle," *Journal of Field Robotics,* vol. 30.3, pp. 371-398, 2013.

[25] A. P. Aguiar and J. P. Hespanha, "Position tracking of underactuated vehicles," in *American Control Conference Proceedings of the 2003. Vol. 3. IEEE*, 2003.

[26] K. D. Do, "Practical control of underactuated ships," *Ocean Engineering,* vol. 37.13, pp. 1111-1119, 2010.

[27] Mahini, Farshad and Hashem Ashrafiuon, "Autonomous Surface Vessel Target Tracking Experiments in Simulated Rough Sea Conditions," in *ASME 2012 5th Annual Dynamic Systems and Control Conference joint with the JSME 2012 11th Motion and Vibration Conference*, 2012.

[28] J. Alvarez, I. R. Bertaska and K. D. von Ellenrieder, "Nonlinear Adaptive Control of an Amphibious Vehicle," in *Proc. ASME Dynamic Systems Control Conference*, 2013.

[29] H. Ashrafiuon, R. M. Kenneth and L. C. McNinch, "Review of nonlinear tracking and setpoint control approaches for autonomous underactuated marine vehicles," in *American Control Conference (ACC), IEEE*, 2010.

[30] A. P. Aguiar and A. M. Pascoal, "Dynamic positioning and way-point tracking of underactuated AUVs in the presence of ocean currents," *International Journal of Control,* vol. 80.7, pp. 1092-1108, 2007.

[31] Pereira, Arvind, D. Jnaneshwar and G. Sukhatme, "An experimental study of station-keeping on an underactuated ASV," in *Intelligent Robots and Systems, 2008. IROS 2008. IEEE/RSJ International Conference on. IEEE*, 2008.

[32] E. Chen, S.-W. Huang, Y.-C. Lin and J.-H. Guo, "Station-keeping of Autonomous Surface Vehicle in Surf Zone," in *MTS/IEEE Oceans*, Bergen, 2013.

[33] G. Elkaim and R. Kelbley, "Station-keeping and Segment Trajectory Control of a Wind—Propelled Autonomous Catamaran," in *45th IEEE Conference on Decision & Control*, San Diego, 2006.

[34] Panagou, Dimitra and K. J. Kyriakopoulos, "Switching control approach for the robust practical stabilization of a unicycle-like marine vehicle under non-vanishing perturbations," in *Robotics and Automation (ICRA), 2011 IEEE International*

Conference on. IEEE, 2011.

[35] Panagou, Dimitra and K. J. Kyriakopoulos, "Dynamic positioning for an underactuated marine vehicle using hybrid control," *International Journal of Control* , vol. 87.2, pp. 264-280, 2014.

[36] T. D. Nguyen, A. J. Sorenson and S. T. Quek, "Design of hybrid controller for dynamic positioning from calm to extreme sea conditions," *Automatica*, vol. 43, no. 5, pp. 768-785, 2007.

[37] Fossen, I. Thor and T. A. Johansen, "A survey of control allocation methods for ships and underwater vehicles," in *Control and Automation, 14th Mediterranean Conference on. IEEE*, 2006.

[38] W. Li, J. Du, Y. Sun, C. Haiquan, Y. Zhang and J. Song, "Modeling and simulation of marine environmental disturbances for dynamic positioned ship," in *31st Chinese Control Conference (CCC)*, 2012.

[39] R. W. F. Gould, *The estimation of wind loads on ship superstructures,* The Royal Institution of Naval Architects, Monograph, 1982.

[40] R. M. Isherwood, "Wind resistance of merchant ships," *Transportation Royal Institution of Naval Architects,* vol. 114, pp. 327-338, 1972.

[41] The Society of Naval Architects and Marine Engineers, "Nomenclature for Treating the Motion of a Submerged Body through a Fluid," *Technical and Research Bulletin* , pp. 1-15, April 1950.

[42] M. Caccia, M. Bibuli, R. Bono and G. Bruzzone, "Basic Navigation, Guidance and Control on an Unmanned Surface Vehicle," *Journal of Autonomous Robots,* vol. 5, no. 4, pp. 349-365, 2008.

[43] Huang, S. Albert, E. Olson and D. C. Moore, "LCM: Lightweight communications and marshalling," in *Intelligent robots and systems (IROS), 2010 IEEE/RSJ international conference on. IEEE*, 2010.

[44] American Bureau of Shipping, *Guide for Vessel Maneuverability,* Houston, 2006, pp. 25-30.

[45] W. B. Klinger, I. R. Bertaska and K. D. von Ellenrieder, "Experimental testing of an adaptive controller for USVs with uncertain displacement and drag," in *MTS/IEEE Oceans*, St. John's, 2014.

[46] N. Mišković, Z. Vukić, M. Bibuli, G. Bruzzone and M. Caccia, "Fast in-field identification of unmanned marine vehicles.," *Journal of Field Robotics,* vol. 28, no. 1, pp. 101-120, 2011.

[47] Y. H. Eng, K. M. Teo, M. Chitre and K. M. Ng, "Online System Identification of an Autonomous Underwater Vehicle Via In-Field Experiments.," *IEEE Journal of Oceanic Engineering,* vol. 41, no. 1, pp. 5-17, 2016.

[48] H. G. Sage, M. F. De Mathelin and E. Ostertag, "Robust Control of robot manipulators: a survey," *Int. J. Control,* 1999.

[49] J.-J. Slotine and W. Li, Applied nonlinear control, Prentice-Hall, 1991, pp. 301-303.

[50] O. M. Faltinsen, Hydrodynamics of High-Speed Marine Vehicles, New York, NY: Cambridge University Press, 2005, p. 85.

[51] Ø. K. Kjerstad and R. Skjetne, "Modeling and Control for Dynamic Positioned Marine Vessels in Drifting Managed Sea Ice," *Modeling, Identification and Control ISSN 1890–1328,* vol. 35, no. 4, p. 249–262, 2014.

[52] Ø. K. Kjerstad, R. Skjetne and B. O. Berge, "Constrained nullspace-based thrust allocation for heading prioritized stationkeeping of offshore vessels in ice," in *Proc. Int. Conf. Port Ocean Eng. Arctic Conditions*, Espoo, Finland, 2013.

[53] Ø. K. Kjerstad and B. M. , "Weather Optimal Positioning Control for Marine Surface Vessels," in *8th IFAC Conference on Control Applications in Marine Systems*, Rostock-Warnemünde, Germany, 2010.

[54] M. H. Zhang, in *Wind Resource Assessment and Micro-siting: Science and Engineering*, John Wiley & Sons, 2015, 2015.

[55] OCIMF, "Prediction of Wind and Current Loads on VLCCs," in *Oil Companies International Marine Forum*, London, 1977.

[56] W. T. P. a. D. A. Van Berlekom, "Large Tankers: Wind Coefficients and Speed Loss Due to Wind and Sea," in *Royal Institution of Naval Architects*, London, 1974.

[57] J. O. De Kat and J. W. Wichers, "Behavior of a Moored Ship in Unsteady Current, Wind and Waves," *Marine Techonology,* vol. 28, pp. 251-264, 1991.

[58] S. Damitha and K. Nihal, "Modeling and Simulation of Environmental Disturbances for Six degrees of Freedom Ocean Surface Vehicle," *Sri Lankan Journal of Physics,* vol. 10.4038/sljp.v10i0.3834, no. Department of Physics, University of Colombo, Colombo 03, 2011.

[59] T. I. Fossen, Guidance and Control of Ocean Vehicles, Chichester: John Wiley and Sons Ltd, 1994.

[60] Webster, C. William and J. Sousa, "Optimum allocation for multiple thrusters," in *Proc. of the Int. Society of Offshore and Polar Engineers Conference (ISOPE'99)*, 1999.

[61] Johansen, Tor Arne, T. I. Fossen and S. P. Berge, "Constrained nonlinear control allocation with singularity avoidance using sequential quadratic programming," *Control Systems Technology, IEEE Transactions* , vol. 12.1, pp. 211-216, 2004.

[62] Yadav, Parikshit and et.al., "Optimal Thrust Allocation for Semisubmersible Oil Rig Platforms Using Improved Harmony Search Algorithm," *Oceanic Engineering,* vol. 39.3, pp. 526-539, 2014.

[63] O. J. Sordalen, "Optimal thrust allocation for marine vessels," *Control Engineering Practice,* vol. 5.9, pp. 1223-1231, 1997.

[64] S. Yu, X. Yu, B. Shirinzadeh and Z. Man, "Continuous finite-time control for robotic manipulators with terminal sliding mode," *Automatica 41,* pp. 1957-1964, 2005.